\documentclass[prx, superscriptaddress, twocolumn, showpacs, amsmath, amssymb, longbibliography]{revtex4-1}

\usepackage{graphicx}
\usepackage{color}
\usepackage{amsmath}
\usepackage{layouts}
\usepackage{siunitx}
\usepackage[colorinlistoftodos]{todonotes}
\usepackage[normalem]{ulem}

\newcommand{\dd}{\text{d}}

\begin{document}

\title{Experimental  {evidence}  of radiation reaction in the collision of a high-intensity laser pulse with a laser-wakefield accelerated electron beam}
%%%%%%%%%%%%%%
% Authors
%%%%%%%%%%%%%%%%%%%%%%%%%%%%%%%%%%%%%%%%%%%%%%

\newcommand{\JAI}
{The John Adams Institute for Accelerator Science, Imperial College London, London, SW7 2AZ, UK}
\newcommand{\GOLP}
{GoLP/Instituto de Plasmas e Fus\~{a}o Nuclear, Instituto Superior T\'{e}cnico, U.L., Lisboa 1049-001, Portugal}
\newcommand{\CUOS}
{Center for Ultrafast Optical Science, University of Michigan, Ann Arbor, Michigan 48109-2099, USA}
\newcommand{\YORK}
{York Plasma Institute, Department of Physics, University of York, York, YO10 5DD}
\newcommand{\QUEENS}
{School of Mathematics and Physics, The Queen's University of Belfast, BT7 1NN, Belfast, UK}
\newcommand{\JENA}
{Institut f\"{u}r Optik und Quantenelektronik, Friedrich-Schiller-Universit\"{a}t, 07743 Jena, Germany}
\newcommand{\STRATH}
{SUPA Department of Physics, University of Strathclyde, Glasgow G4 0NG, UK}
\newcommand{\CHALMERS}
{Department of Physics, Chalmers University of Technology, SE-41296 Gothenburg, Sweden}
\newcommand{\CLF}
{Central Laser Facility, Rutherford Appleton Laboratory, Didcot OX11 0QX, UK}
\newcommand{\PLYMOUTH}
{Centre for Mathematical Sciences, Plymouth University, UK}
\newcommand{\UCLA}
{University of California, Los Angeles, Los Angeles, California, 90095, USA}
\newcommand{\LANCS}
{Physics Department, Lancaster University, Bailrigg, Lancaster LA1 4YW, UK}

\author{J.~M.~Cole}
\email{j.cole11@imperial.ac.uk}
\affiliation{\JAI}
\author{K.~T.~Behm}
\affiliation{\CUOS}
\author{E.~Gerstmayr}
\affiliation{\JAI}
\author{T.~G.~Blackburn}
\affiliation{\CHALMERS}
\author{J.~C.~Wood}
\affiliation{\JAI}
\author{C. D.~Baird}
\affiliation{\YORK}
\author{M.~J.~Duff}
\affiliation{\STRATH}
\author{C.~Harvey}
\affiliation{\CHALMERS}
\author{A.~Ilderton}
\affiliation{\CHALMERS}
\affiliation{\PLYMOUTH}
\author{A.~S.~Joglekar}
\affiliation{\UCLA}
\author{K. Krushelnick}
\affiliation{\CUOS}
\author{S.~Kuschel}
\affiliation{\JENA}
\author{M.~Marklund}
\affiliation{\CHALMERS}
\author{P.~McKenna}
\affiliation{\STRATH}
\author{C.~D.~Murphy}
\affiliation{\YORK}
\author{K.~Poder}
\affiliation{\JAI}
\author{C.~P.~Ridgers}
\affiliation{\YORK}
\author{G.~M.~Samarin}
\affiliation{\QUEENS}
\author{G.~Sarri}
\affiliation{\QUEENS}
\author{D.~R.~Symes}
\affiliation{\CLF}
\author{A.~G.~R.~Thomas}
\affiliation{\CUOS}
\affiliation{\LANCS}
\author{J.~Warwick}
\affiliation{\QUEENS}
\author{M.~Zepf}
\affiliation{\JENA}
\affiliation{\QUEENS}
\author{Z.~Najmudin}
\affiliation{\JAI}
\author{S.~P.~D.~Mangles}
\email{stuart.mangles@imperial.ac.uk}
\affiliation{\JAI}

\date{\today}

%%%%%%%%%%%%%%%%%%%%%%%%%%%%%%%%%%%%%%%%%%%%%%
\begin{abstract}
%%%%%%%%%%%%%%%%%%%%%%%%%%%%%%%%%%%%%%%%%%%%%%

The dynamics of energetic particles in strong electromagnetic fields can be heavily influenced by the energy loss arising from the emission of radiation during acceleration, known as radiation reaction. 
When interacting with a high-energy electron beam, today's lasers are sufficiently intense to explore the transition between the classical and quantum radiation reaction regimes.
 {We present evidence of radiation reaction} in the collision of an ultra-relativistic electron beam generated by laser wakefield acceleration ($\varepsilon > 500$~MeV) with an intense laser pulse ($a_0 > 10$). 
We measure an energy loss in the post-collision electron spectrum that is correlated with the detected signal of hard  photons ($\gamma$-rays), consistent with a  {quantum} description of radiation reaction. 
The generated $\gamma$-rays have the highest energies yet reported from an all-optical inverse Compton scattering scheme, with critical energy $\varepsilon_{\rm crit} > $~30~MeV.
\end{abstract}

\maketitle

%%%%%%%%%%%%%%%%%%%%%%%%%%%%%%%%%%%%%%%%%%%%%%
\section{Introduction}
%%%%%%%%%%%%%%%%%%%%%%%%%%%%%%%%%%%%%%%%%%%%%%

Accelerating charges radiate and therefore lose energy. 
The effective force on charged particles resulting from these losses, known as radiation reaction (RR), scales quadratically with both particle energy and applied electromagnetic field strength. 
Normally radiation reaction is negligible but it becomes comparable in magnitude to the Lorentz force on an electron when $\gamma E$ approaches $E_{cr}$, where $E$ is the electric field on a particle of Lorentz factor $\gamma$, and $E_{cr} = m_e^2c^3/\hbar e = 1.3\times10^{18}$~Vm$^{-1}$ is the critical field of quantum electrodynamics (QED).
High electric fields and electron energies are then required to observe radiation reaction, a regime which may occur in astrophysical contexts~\cite{Tchufistova2011, Jaroschek2009} and the laser-plasma interaction physics that will be explored at next-generation, 10~PW class laser facilities~\cite{DiPiazza2012a, Thomas2012}.
In the weak field classical limit there are different formulations of radiation reaction~\cite{Hammond2010,Burton2014b}; the most widely used is that of Landau and Lifshitz (LL)~\cite{Landau1975} which can be derived from the low-energy limit of QED~\cite{Krivitskii1991,Ilderton2013}.
A notable deficiency of classical models is that the radiation spectrum is unbounded, allowing the emission of photons with more energy than the electron.
Classical models therefore over-estimate radiation reaction forces and emitted photon energies compared to quantum-corrected models~\cite{Burton2014b,Neitz2013,Yoffe2015,Vranic2016a,Dinu2016,Krivitskii1991,Ilderton2013}.

The collision of a high-energy electron bunch with a tightly-focussed, intense laser pulse provides a suitable configuration for the observation of radiation reaction.
Experimentally realising the high intensities required for this necessitates the use of laser pulses of femtosecond duration, and so synchronisation between the electron bunch and the colliding laser pulse must also be maintained at the femtosecond level.
Laser-wakefield accelerators are plasma-based electron accelerators driven by intense laser pulses~\cite{Mangles2004a,Geddes2004,Faure2004,Esarey2009a}, capable of accelerating electron beams to the GeV level  {\cite{Leemans2006, Kneip2009,  Leemans2014, Wang2013}}.
The high electron beam energy coupled to the intrinsic synchronisation with the driving laser pulse means that wakefield accelerators are uniquely suited to the study of ultrafast laser-electron beam interactions, and have been the focus of much recent work~\cite{TaPhuoc2012a,Chen2013a,Sarri2014,Yan2017}. 
In our scheme, one laser pulse is used to drive a wakefield accelerator while a second, counter-propagating pulse collides with the electron bunch.
The electrons oscillate in the fields of the second laser and back-scatter radiation boosted in the direction of the bunch, a process known as inverse Compton scattering (ICS).

The spectrum of the scattered photons is determined by the normalised laser amplitude $a_0 = 0.855\lambda_0[\si{\um}]I^{1/2}[10^{18}\text{Wcm}^{-2}]$, the laser frequency $\omega_0 = 2\pi c /\lambda_0$, and the electron beam energy.
In the low $a_0$ limit the electron motion is simple harmonic and the back-scattered photon energy is the Doppler-upshifted laser photon energy $\hbar\omega = \hbar\omega_0\gamma(1+\beta)/(\gamma(1-\beta) + 2\hbar\omega_0/m_ec^2) \simeq 4\gamma^2\hbar\omega_0$ for $\gamma \gg 1$ and $\hbar\omega_0 \ll m_ec^2$.
All-optical experimental configurations involving the collision of wakefield accelerated electron beams with laser pulses in this regime have produced scattered x-rays with energies in the range of hundreds of keV~  {\cite{TaPhuoc2012a, Powers2014, Tsai2015}}.

As $a_0$ increases, the scattered photon energy initially decreases as $\hbar\omega\simeq 4\gamma^2\hbar\omega_0/(1 + a_0^2/2)$, measured experimentally for $a_0 < 1$~\cite{Sakai2015, Khrennikov2015}.
The electron motion becomes anharmonic and it begins to radiate higher harmonics, or equivalently interacts with multiple photons in the nonlinear regime of Compton scattering~\cite{Bula1996,Sarri2014,Yan2017}.
For $a_0 \gg 1$ the effective harmonic order increases as $a_0^3$ and the spectrum of the scattered radiation assumes a broad synchrotron-like form.
The characteristic energy of the spectrum $\varepsilon_{\text{ICS}} = 3\gamma^2a_0\hbar\omega_0$~\cite{Esarey1993b} increases with $a_0$.
The fraction of the electron energy lost per photon emission is then of order $\varepsilon_{\text{ICS}}/\gamma m_ec^2 = 3\eta/2$ where $\eta = 2\gamma a_0 \hbar\omega_0/m_ec^2$ is the quantum nonlinearity parameter in this geometry~\cite{Ritus1985a}, the ratio of the laser electric field to $E_{cr}$ in the rest frame of the electron. 
Strong field quantum effects are present even when $\eta\ll1$~~\cite{Nerush2011,Thomas2012}; as $\eta$ approaches unity the impact of radiation reaction on the electron and discrete nature of the photon emission cannot be neglected when calculating the photon spectrum~\cite{Neitz2013, Blackburn2014}, and the scaling of $\varepsilon_\text{ICS}$ with $\gamma$ and $a_0$ slows.
This is known as the quantum regime of radiation reaction. 

Here we describe an experiment which probes radiation reaction by simultaneously measuring the electron and Compton-scattered photon spectra after the collision of a wakefield accelerated electron beam with an intense laser pulse.
We observe scattered $\gamma$-rays at the highest energies measured to date in a wakefield-driven inverse Compton scattering experiment.
Independent measurements of the $\gamma$-ray spectrum and the electron energy after the collision are only consistent when radiation reaction is taken into account, and we find that the internal consistency of these measurements is improved when a fully quantum (stochastic) description of radiation reaction is used.  

%%%%%%%%%%%%%%%%%%%%%%%%%%%%%%%%%%%%%%%%%%%%%%
\section{Experiment}
%%%%%%%%%%%%%%%%%%%%%%%%%%%%%%%%%%%%%%%%%%%%%%

\begin{figure}
	\includegraphics[width=\columnwidth]{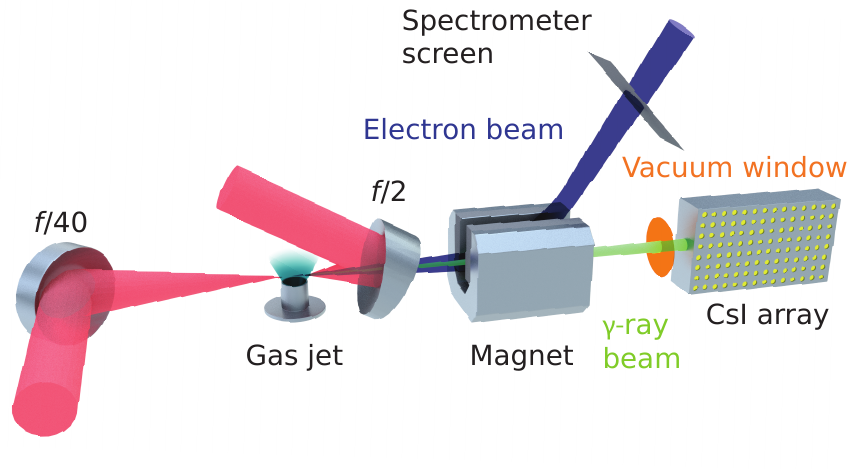}
	\caption{
		Schematic of the experimental setup. All components are inside a vacuum chamber except for the CsI array.
	}
	\label{fig:setup}
\end{figure}

The experiment was conducted using the Astra-Gemini laser of the Central Laser Facility, Rutherford Appleton Laboratory, UK.
Figure~\ref{fig:setup} shows a schematic of the experimental setup.
Gemini is a Ti:Sapphire laser system delivering two synchronised linearly polarised beams of 800\,nm central wavelength and pulse durations of 45\,fs full-width-half-maximum (\textsc{fwhm}).
One of the beams, used to drive a laser wakefield accelerator, was focussed with an $f$/40 spherical mirror to a focal spot \textsc{fwhm} size of $37\,\times\,49$\,\si{\um}.
The energy delivered to the target was $(8.6\,\pm\,0.6)$\,J generating a peak intensity of $(7.7\,\pm\,0.4)\,\times\,10^{18}$\,Wcm$^{-2}$, corresponding to a peak normalised amplitude of $a_0 = 1.9\,\pm\,0.1$.
This pulse was focussed at the leading edge of a 15\,mm-diameter supersonic helium gas jet, which produced an approximately trapezoidal density profile with 1.5\,mm linear ramps at the leading and trailing edges.
Once ionised by the laser, the peak plasma electron densities used here were $(3.7\,\pm\,0.4)\,\times\,10^{18}$\,cm$^{-3}$.

The second Gemini beam was focussed at the rear edge of the gas jet, counter-propagating with respect to the first.
As the laser-wakefield generated electron beam interacted with the second focussed laser pulse, inverse Compton scattered $\gamma$-rays were generated, co-propagating with the electron beam.
 {By colliding close to the rear of the gas jet, the electron bunch did not have time to diverge before the collision and so the overlap between the electron bunch and laser was maximised.}

The focussing optic for the second pulse was an off-axis $f$/2 parabolic mirror with a hole at the centre to allow free passage of the $f$/40 beam, electron beam, and scattered $\gamma$-rays. 
Accounting for the hole in the optic, the pulse energy on target was $(10.0\,\pm\,0.6)$\,J.
This was focussed to a focal spot \textsc{fwhm} size of $2.4\,\times\,2.8$\,\si{\um} at a peak intensity of $(1.3~\pm~0.1)~\times~10^{21}$~Wcm$^{-2}$, corresponding to a peak normalised amplitude $a_0 = 24.7\,\pm\,0.7$.

In order to align the two laser beams onto the same optical axis, a 90$^{\circ}$ prism with a micrometre-sharp edge was inserted into the beamline at the interaction point.
After overlapping the focussed pulses on the tip of the prism, half of each was reflected collinearly onto a CCD~\cite{Faure2006} and  imaged with a 10$\times$ magnification microscope objective. 
After reflection the different wavefront curvatures of the $f$/2 and $f$/40 beams caused circular interference fringes to appear when the pulses overlapped in time.
By optimising the fringe visibility the two pulses were overlapped to a precision of $\pm\,30$~fs, limited by the random optical path length fluctuations during the measurement.

\begin{figure}
	\includegraphics[width=\columnwidth]{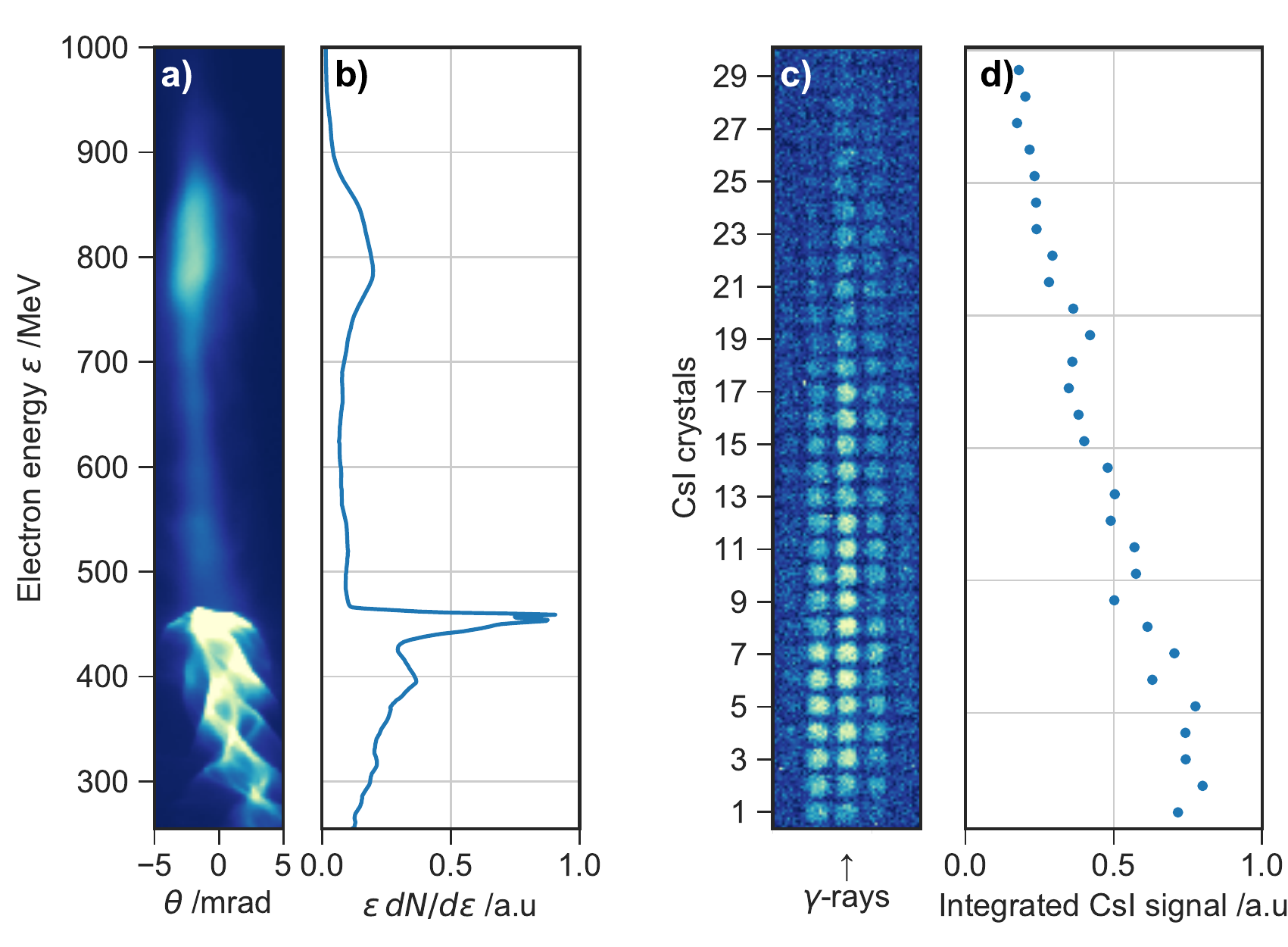}
	\caption{
		\textbf{a)} Electron spectrometer screen image, transformed onto a linear energy axis.
		\textbf{b)} Angularly-integrated electron spectrum. 
		\textbf{c)} Raw image of CsI crystal stack detector. 
		\textbf{d)} Integrated CsI signal as a function of penetration depth into the stack.
	}
	\label{fig:rawdata}
\end{figure}

After passing through the hole in the $f$/2 mirror, the electron beam was deflected from the optical axis by a permanent dipole magnet with total magnetic length $\int B(x)\,\dd x$ = 0.4~Tm.
The electron energy spectrum was recorded in the range of 0.25 -- 2~GeV on a scintillating Gd$_2$O$_2$S:Tb (Lanex) screen placed in the path of the magnetically-dispersed beam, and imaged with a cooled 16-bit CCD camera. An exemplary spectrometer image is shown in Fig.~\ref{fig:rawdata} a), and the calculated electron spectrum in Fig.~\ref{fig:rawdata} b).

The $f$/40 laser pulse was blocked at the rear of the interaction chamber with a 50~\si{\um}-thick aluminium foil, which along with a 250~\si{\um}-thick Kapton vacuum window was traversed by the $\gamma$-ray beam. 
The $\gamma$-ray detector consisted of an array of $5\,\times\,5\,\times\,50$~mm caesium iodide (CsI) crystals doped with thallium, which convert deposited energy into optical photons at an efficiency of $\approx 5\,\times\,10^4$~MeV$^{-1}$.
The array was 33 crystals high and 47 crystals in the longitudinal direction, with the $\gamma$-rays entering through the $5\,\times\,50$~mm faces.
The crystals were separated by 1~mm thick aluminium spacers, and the face of the stack exposed to the $\gamma$ beam was covered with a 9~mm thick stainless steel plate.
By imaging the $5\,\times\,5$~mm faces of the CsI crystals from the side and recording the scintillation light, it was possible to record a vertically-resolved map of the energy deposition in the detector -- see Fig.~\ref{fig:rawdata} b).
Low-energy photons deposit most of their energy in the first crystal column, with the energy deposited in subsequent crystals decreasing monotonically.
High-energy photons create an electromagnetic shower which causes the energy deposition to initially increase with depth before decaying.

%%%%%%%%%%%%%%%%%%%%%%%%%%%%%%%%%%%%%%%%%%%%%%%%%%
\section{Electron spectra and $\gamma$-ray yield}
\label{sec:ElectronSpectra}
%%%%%%%%%%%%%%%%%%%%%%%%%%%%%%%%%%%%%%%%%%%%%%%%%%

\begin{figure}
	\includegraphics[width=\columnwidth]{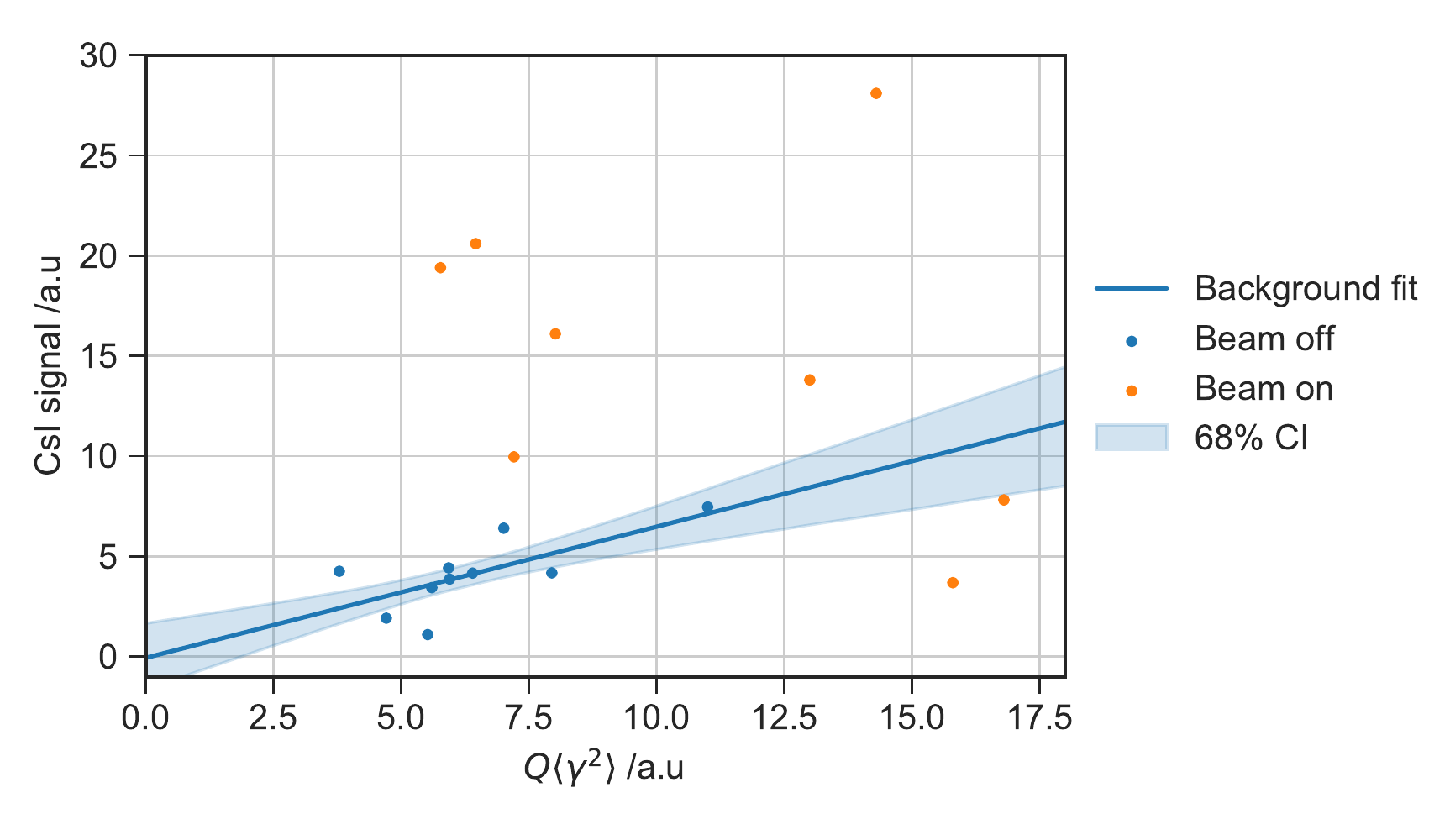}
	\caption{
The total CsI signal as a function of the integrated squared energy of the electron spectrum, and the linear fit to the `beam-off' shots. 
The shaded area represents the 68\% confidence interval (CI) for the linear fit.
	}
	\label{fig:csivsqg2}
\end{figure}

The data analysed here is a sequence of 18 shots where electron spectra and $\gamma$-ray signals were recorded simultaneously.
For the first 8 shots the $f$/2 beam was on, then for the next 10 shots it was switched off.
Due to the shot to shot variations of the electron beam and laser pointing and timing we do not expect every collision to be successful (based on the measured fluctuations we expect approximately 1 shot in 3 to occur at an $a_0$ that is large enough to produce a measurable radiation reaction).  
Before we can proceed in assessing if radiation reaction was occuring in our experiment, we must therefore first  identify which collisions were successful. 
This can be achieved by analysing the $\gamma$-ray signal -- successful collisions will produce a much brighter signal than those which are not.
In Fig.~\ref{fig:csivsqg2} we plot the integrated signal on the CsI detector as a function of the electron beam properties.

It is important to account for any background which could contaminate the CsI detector (such as bremsstrahlung emission), as this will increase with electron beam charge and energy in the same way as the inverse Compton signal. 
Specifically, for an electron of Lorentz factor $\gamma$ the total energy of the emitted radiation scales as $\gamma^2$. 
An electron spectrum $\dd N_e/\dd\gamma$ will therefore create background signal with an energy proportional to $\int \gamma^2 (\dd N_e/\dd\gamma)\,\dd\gamma = Q\langle\gamma^2\rangle$ where $Q$ is the total beam charge. 
The energy radiated into the ICS beam, and therefore the integrated signal on the CsI detector, is approximately proportional to $a_0^2Q\langle\gamma^2\rangle$ for $\gamma a_0^2 < 5.5\times10^5$ \cite{Thomas2012}, and so the total signal is expected to be

\begin{equation}
\text{CsI signal} = c_{\rm BG}Q\langle\gamma^2\rangle + c_{\rm ICS}a_0^2Q\langle\gamma^2\rangle
\label{eqn:signalmodel}
\end{equation}

\noindent for some constants $c_{\rm BG}$, $c_{\rm ICS}$.
In Fig.~\ref{fig:csivsqg2} a linear estimation of $c_{\rm BG}$ is performed using the `beam-off' shots, providing an estimate (with error) of the background signal in the `beam-on' shots.

\begin{figure*}
	\includegraphics[width=\textwidth]{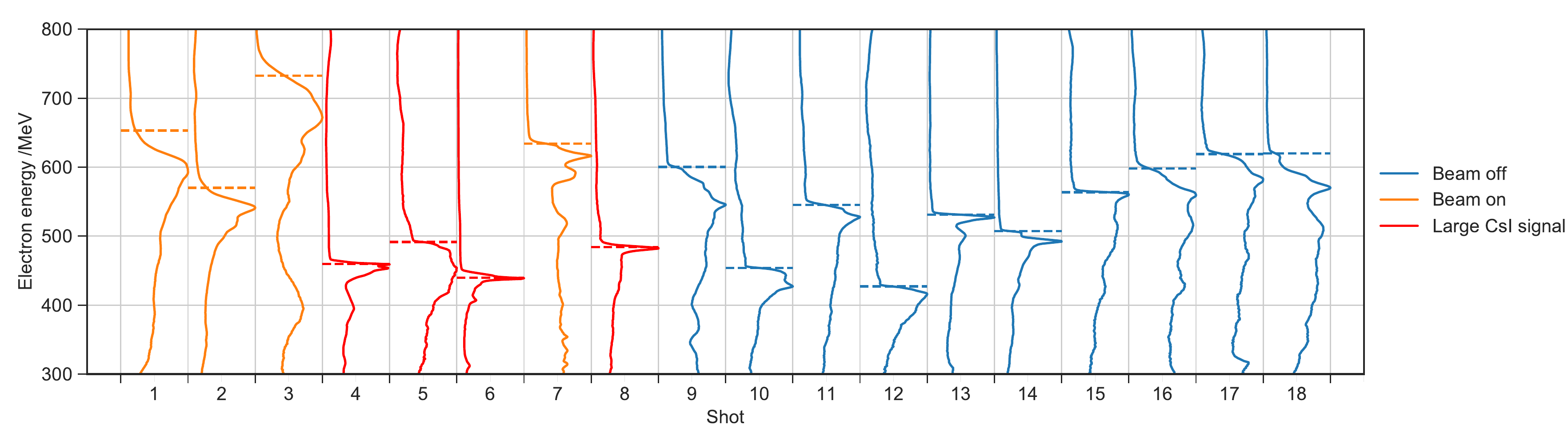}
	\caption{
			Consecutive electron spectra. 
			Each spectrum has been normalised to its own maximum.
	}
	\label{fig:electronspectra}
\end{figure*}

The consecutive angularly-integrated electron spectra are plotted in Fig.~\ref{fig:electronspectra}.
The spectra were almost always observed to possess two components -- a high-charge, low-energy feature, and a low-charge, high-energy feature. 
It is possible that this is due to separate injection events caused by density structures in the plasma, observable on transverse plasma diagnostics and likely due to fluid shocks in the gas flow. 
This would imply that the high energy component was generated by self-injection~\cite{Bulanov1997} and that the low energy component was injected at an abrupt density transition~\cite{Schmid2010}.
In Fig.~\ref{fig:electronspectra} the energy at which these features become distinct is highlighted, overlaid on the electron spectrum.
We refer to this feature as an `edge' in the spectrum.

\begin{figure}
\includegraphics[width=\columnwidth]{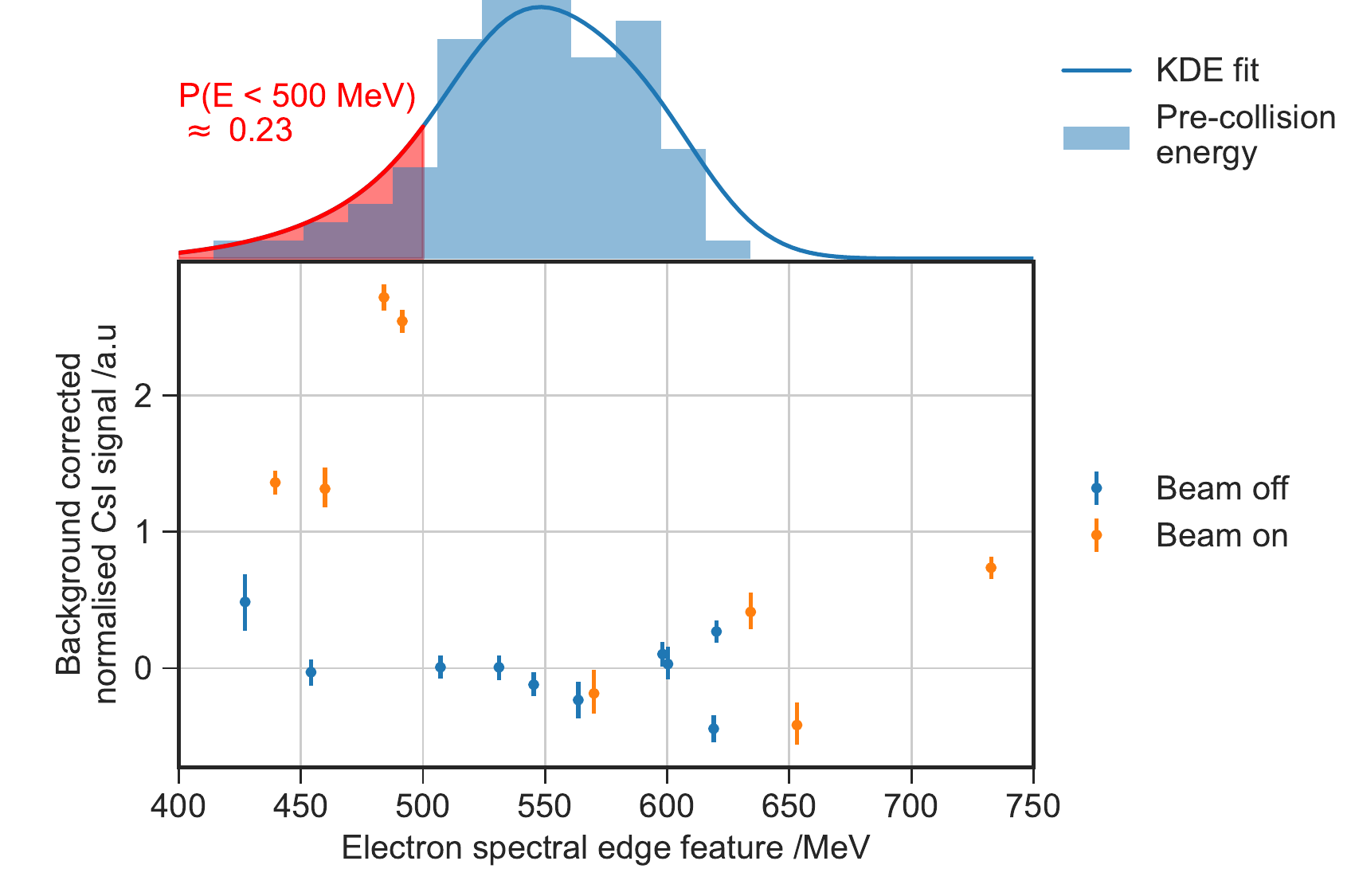}
\caption{
  The background-subtracted CsI signal as a function of the energy of the edge in the electron spectrum.
  The histogram above shows the distribution of electron beam energies recorded during a separate sequence of shots where no radiation reaction signal could occur.
  Overlaid is a kernel density estimate (KDE) to the distribution of electron beam energies. 
	}
	\label{fig:csivsef}
\end{figure}

The quantity determining the magnitude of the interaction is, in the terms of Eq.~\ref{eqn:signalmodel}, $c_{\rm ICS}a_0^2$.
This a factor representing the overlap of the electron beam and focal spot, and so the effective $a_0$ of the interaction.
The quantity of interest is then calculated from the measured signal by subtracting an estimated background, and dividing by $Q\langle\gamma^2\rangle$.
This is the quantity plotted in Fig.~\ref{fig:csivsef} as a function of the measured energy of the edge in the electron spectrum.

This corrected signal for the `beam-off' shots is clustered around zero, as expected, but also for some of the `beam-on' shots, as expected due to the spatio-temporal jitter between the laser focal spot and the electron beam.
There are however four shots with exceptionally large signals, more than four standard deviations above the background level -- shots 4, 5, 6, and 8.
The chance of observing a signal this large in a sample of 8 shots due to background fluctuations alone is less than 0.1\%.
These shots are highlighted in red in Fig.~\ref{fig:electronspectra}.

Using this signal threshold as an independent criterion for assessing which shots successfully resulted in a strong laser-electron beam interaction, we then consider the electron energies for these collisions.
All four electron beam energies are below 500~MeV, while the mean of the 10 beam-off shots was measured to be $(550 \pm 20)$~MeV, with a standard deviation of $(63 \pm 14)$~MeV.
Assuming that the electron energies on the shots with the f/2 beam off are sampled from a distribution which is approximately normal, this implies that there would be a $\sim$23\% chance of observing an electron energy below 500~MeV on a single successful collision even if the interaction did not result in energy loss.  
However, the chance of observing an energy less than 500~MeV on all four successful collision shots falls to 0.3\% under this null hypothesis. 
This is sufficiently unlikely that we conclude that the low energies observed on the four successful collision shots are due to the radiation reaction force on the electron bunch.
This analysis assumes that the beam-off shots are sampled from a normal distribution. 
To assess this a larger set of 87 shots where no radiation reaction signal could occur was recorded under similar conditions.
The distribution of energies from this larger data set is shown as a histogram in Fig.~\ref{fig:csivsef}.
An Anderson-Darling test confirms that the 10 beam off shots are consistent with being sampled from a normal distribution, at the 99.9\% confidence level.
Note also that, under the null hypothesis, the chance of observing four or more electron beams with energies below 500~MeV in a sample of 8 is much higher, approximately 10\%, and so taken alone the electron beam energies would not have represented a clear observation.
It is only by independently identifying which collisions were successful, based on the observation of a bright $\gamma$-ray signal, that the observed low energies become statistically significant.

From the observed decrease in electron energy in the shots generating a bright CsI signal, we can estimate the laser intensity required to cause this energy shift under different radiation reaction models.
Assuming that the `bright' shots and `beam-off' shots are sampled from two different normal distributions with different mean values, we consider the energy shift between these mean values from $\varepsilon_{\text{initial}} = (550\,\pm\,20)$~MeV to $\varepsilon_{\text{final}}=(470\,\pm\,10)$~MeV. 
If the electron beam interacts with a pulse of \textsc{fwhm} duration 45~fs, the required peak $a_0$ to generate this energy loss is 10$\,\pm\,$2 for a quantum model, and 9$\,\pm\,$1 for a classical model based on the Landau-Lifshitz equation.
For $\varepsilon_{\text{initial}} \approx $ 550~MeV and $a_0 \approx$ 10, $\eta \approx 0.07$ and in the quantum model the electron loses energy at a rate $\approx 0.75$ that of the classical model~\cite{Ridgers2014}.
In the classical model a slightly lower $a_0$ is therefore able to generate the same electron energy loss, though due to the relatively low value of $\eta$ the difference in $a_0$ between the models is small compared to the experimental uncertainties when considering the electron beam energy loss alone.
Additionally the peak $a_0$ of the $f$/2 focal spot was calculated to be greater than 20 from measured pulse parameters, which is much larger than that inferred from the electron energy shift.

In practice the effective $a_0$ of the interaction should be expected to be significantly lower than 20, due to the finite size of the electron beam and any timing offset of the collision point, since this will result in the interaction occurring a distance from the laser focus.
Post-experiment we identified a systematic offset caused by the delay between the $f$/40 laser pulse and the wakefield accelerated electron bunch.
During the alignment of the experiment the two pulses were temporally overlapped at the rear edge of the gas jet, but this is not the same point as the collision between the electron beam and the $f$/2 pulse.
This is because the wakefield electron bunch will be trailing the $f$/40 pulse by approximately half a plasma wavelength, and therefore the location of the collision between the electron bunch and the $f$/2 beam will be offset from this position by

\begin{equation}
\delta z = \frac{3d}{4}\frac{n_e}{n_c} + \frac{\lambda_0}{4}\sqrt{\frac{n_c}{n_e}}
\end{equation}

\noindent where $d$ is the electron injection point (measured from the front of the gas jet), $n_e$ the electron density, $n_c$ the critical density, and $\lambda_0$ the laser wavelength. 
Here it is assumed that the laser travels in the plasma at the non-linear group velocity~\cite{Decker1996a} and that the electrons travel at $c$ from their injection point.
Assuming a uniform distribution for $d$ between 0 and 10~mm and a normal distribution of $\pm\,30$~fs for the timing jitter, the maximum expected interaction  {$a_0$ at the collision  (averaged over an area of $10\,\si{\um}^2$) is 12$\,\pm\,$1, where the laser transverse size is approximately 5\,\si{\um} \textsc{fwhm}}.
This figure has been corrected for the measured change in size of the focal spot between the low-power alignment modes and the full-power shot mode of 7\%.
This is not the peak $a_0$ of the spot, but the maximum $a_0$ which encloses a contour of area 10~\si{\um}$^2$, an area of similar size to the electron beam plus shot-to-shot position fluctuations of the focal spot.
The variation of $a_0$ near focus under this criterion is slower than the variation of peak $a_0$, and so shot-to-shot timing jitter has less of an impact on the effective interaction $a_0$ than might be expected.
It is very difficult to measure this effective $a_0$, and therefore problematic to distinguish between different radiation reaction models using only the shift in energy of a single feature in the electron spectrum.
While we are confident that we have observed radiation reaction effects, it is not possible from our electron spectral measurements alone to investigate this process in more detail, due to the inherent uncertainty in $a_0$.
To help assess compatibility with different radiation reaction models we therefore augment the electron beam measurements with spectral data from the $\gamma$-ray beam in the following section. 

%%%%%%%%%%%%%%%%%%%%%%%%%%%%%%%%%%%%%%%%%%%%%%
\section{$\gamma$-ray spectra}
%%%%%%%%%%%%%%%%%%%%%%%%%%%%%%%%%%%%%%%%%%%%%%

%---------------------------------------------
\subsection{Measurements}
%---------------------------------------------

We measure the $\gamma$-ray spectrum experimentally by analysing the scintillation yield, and thus energy deposited, in the CsI scintillator array.
To understand the response of the detector, detailed Monte-Carlo modelling of the array was performed in GEANT4~\cite{Agostinelli2003} and MCNP~\cite{Goorley2012} in full 3D, where the simulation geometry included the large objects inside the vacuum chamber, the electron spectrometer magnets, the vacuum chamber itself, and all of the components of the CsI array.
For $\gamma$-ray energies between 2 and 500 MeV, $10^6$ photons were propagated from the electron-laser interaction point into the array. 
The energy deposited in each crystal element was recorded and the scintillation light output was assumed proportional to the deposited energy, as is the case for high-energy photons~\cite{Frlez2000}.
With the detector output as a function of incident photon energy known, it was possible to use a measured detector output to calculate a best-fit $\gamma$-ray spectrum.
A more detailed description of the $\gamma$-ray spectrometer data analysis is presented in reference~\cite{Behm2017}.
From simulations of the inverse Compton scattering process (see below) a good parametrised approximation to $\gamma$-ray spectrum over a wide photon energy range was observed to be

\begin{equation}
\frac{\dd N_{\gamma}}{\dd \varepsilon_{\gamma}} \propto \varepsilon_{\gamma}^{-2/3}e^{-\varepsilon_\gamma/\varepsilon_\text{crit}}
\end{equation}

\noindent where $\varepsilon_\text{crit}$ is a parameter controlling the spectral shape.
For this parametrisation the mean photon energy is $\varepsilon_\text{crit}/3$ and 49\% of the photon energy is radiated below $\varepsilon_\text{crit}$, so $\varepsilon_\text{crit}$ is a characteristic energy of the spectrum.
In the experimental measurements we treat $\varepsilon_{\rm crit}$ as a free parameter and minimise the mean-squared deviation between the simulated and measured detector light yield.
Errors in $\varepsilon_\text{crit}$ were assigned by forming simulated detector response curves and adding synthetic noise at similar levels to that observed in the experimental data, then averaging the retrieved $\varepsilon_{\rm crit}$ over 50 fits.
In this way the 1$\sigma$ fractional fit error was found to be $\pm$~15\%.

\begin{figure}
	\includegraphics[width=\columnwidth]{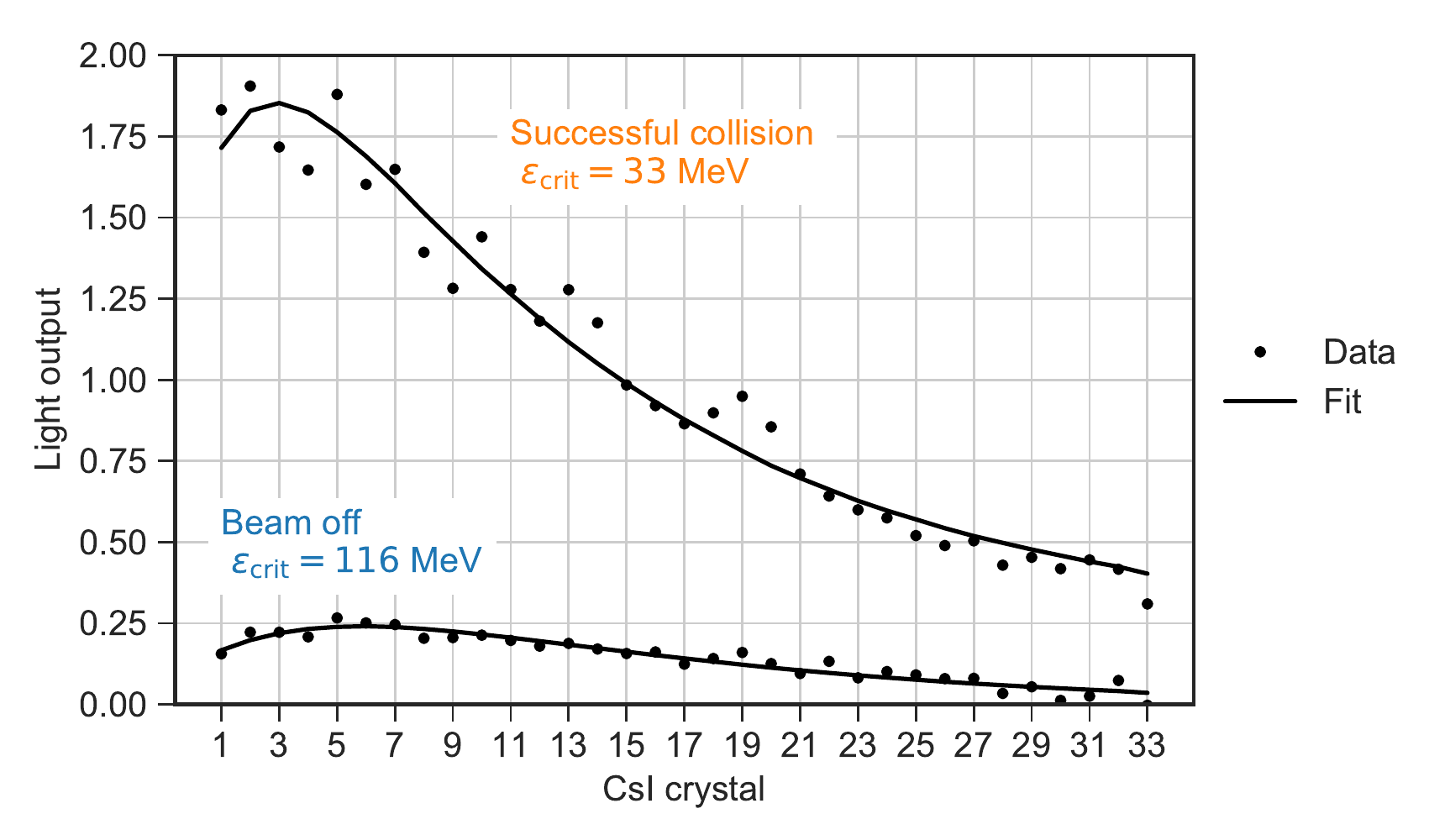}
	\caption{
			CsI spectrometer measurements recorded for a successful collision and `beam-off' shots, and associated best-fit response functions from the Monte-Carlo model of the detector.
			The maximum likelihood estimate of the fit parameter $\varepsilon_{\rm crit}$ is displayed next to each fit.
	}
	\label{fig:csifit}
\end{figure}

	Exemplary fits to data from the $\gamma$-ray detector are shown for a successful collision and a ``beam-off'' shot in Fig.~\ref{fig:csifit}. 
	When bright inverse Compton-scattered $\gamma$-rays are observed, the best-fit value of $\varepsilon_{\rm crit}$ is in the range of several tens of MeV.
	With the colliding beam switched off, the spectrum of the background signal is observed to be much harder, typically with $\varepsilon_{\rm crit} > 100$~MeV.
	This is consistent with a background composed of primarily bremsstrahlung photons produced when the electron beam impacts the walls of the vacuum chamber.

%---------------------------------------------
\subsection{Simulations}
%---------------------------------------------

In order to calculate a theoretical $\gamma$-ray spectrum, Monte-Carlo simulations of the laser-electron collision were performed for different $a_0$ and electron beam energies.
In the simulations quantum and classical models of radiation reaction are compared and contrasted against a control in which no radiation reaction is included.

In the quantum description photon emission is a series of discrete events, the locations of which are stochastically determined based on emission probabilities calculated in the locally constant field approximation  {(LCFA)}~\cite{Ritus1985a}.
 {This is valid for $a_0^3/\eta \gg 1$~\cite{Dinu2016}, as is the case here; its use is necessary because the number of photons emitted per electron is much larger than one, making exact calculation of the spectrum from QED intractable at present. While the LCFA causes the low-energy part of the photon spectrum ($<$ 100 keV here) to be overestimated, these photons do not contribute significantly to the recoil of the electron or to the total radiated energy~\cite{Harvey2015}.}

Between  {individual emission} events the electron follows a classical trajectory with motion determined by the Lorentz force. 
In the standard numerical implementation emission events are determined using Monte Carlo sampling as described in references~\cite{Ridgers2014,Gonoskov2015}. 
This approach has been used to study the production of photons and radiation reaction effects in the experimental configuration considered here~\cite{Bulanov2013,Blackburn2014,Vranic2016a}. 
Equivalently one could solve the kinetic equations for the electron distribution function~\cite{Neitz2013,Sokolov2010,Elkina2011}. 
The approach we use here is based on single particle dynamics in prescribed fields because collective effects are negligible in the considered parameter regime. 
Simulated spectra were obtained and cross-checked using a suite of codes including EPOCH~\cite{Arber2015}, SIMLA~\cite{Green2015} and that used in reference~\cite{Blackburn2015}, which confirmed that collective effects were negligible.

Turning to the classical description, the electron trajectory is determined by integrating an equation of motion which includes both the Lorentz force and energy loss as described by the Landau-Lifshitz radiation reaction force. 
For simplicity we take only the dominant term from this force; as the next term is a factor $2\gamma^2$ smaller, and across the parameter regime we consider $1/(2\gamma^2) \ll 1$, this approximation is appropriate.

While there are other classical models of radiation reaction we do not expect there to be any differences between them for the energy and intensity parameters considered here~\cite{Burton2014b,Vranic2016}.
The emitted $\gamma$-ray spectrum is obtained by sampling the classical synchrotron distribution. 
We expect this model to over-predict the energy loss, as classical descriptions of radiation reaction unphysically fail to preclude the emission of photons with energy higher than that of the seed electron~\cite{Yoffe2015}.

Finally in the `No RR', or control, model, photon emission is calculated as in the quantum case above (to ensure that photons cannot be emitted with an energy larger than the electron energy), but the recoil of the emitting particle is neglected.
This control case will be used to show that neglecting radiation reaction is incompatible with the experimental results obtained, indicating that a regime where radiation reaction is important has been reached (in contrast to previous experiments of this type~\cite{Sarri2014, Yan2017}).

 {One could also consider a modified classical model~\cite{Sokolov2010}, in which the energy loss is continuous but scaled by a  correction factor, $g(\eta)$, that emerges from the quantum theory of photon emission in constant crossed fields. 
However, since the stochastic QED model described above and the modified classical model have been shown to give the same average behaviour ~\cite{Ridgers2017}, and as our experimental data is effectively a measure of the mean energy loss, there would be no significant difference between the predictions of a modified classical model and the fully stochastic calculation we have used at the values of $\eta$ reached in this experiment.
Our simulations indicate that signatures of stochastic emission will be become apparent at $\eta \approx 0.25$.}

\begin{figure}
	\includegraphics[width=\columnwidth]{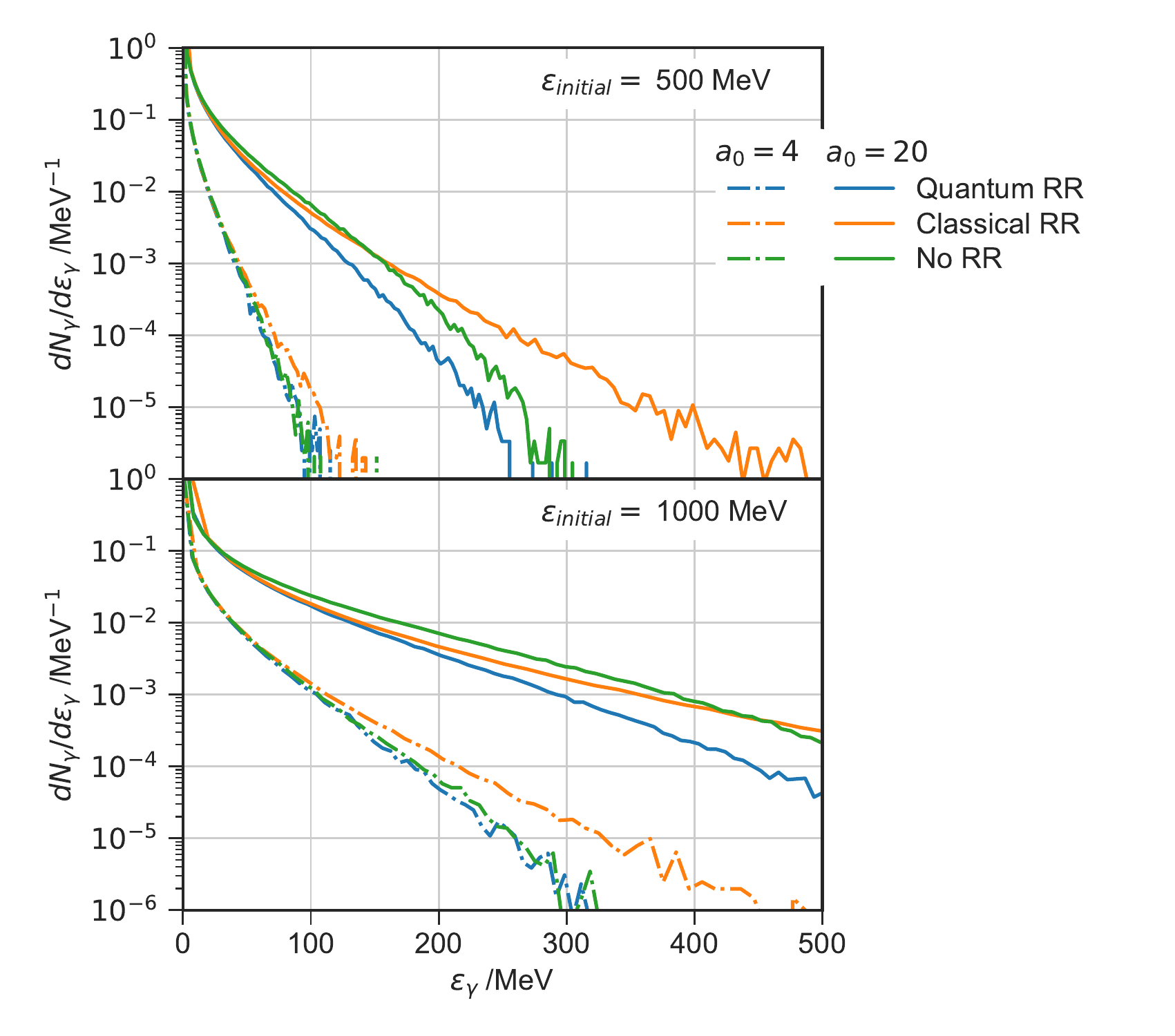}
	\caption{
		The spectrum of ICS photons radiated by a single electron as simulated for various $\varepsilon_{\text{initial}}$ and $a_0$ for the three radiation reaction models described in the text.
	}
	\label{fig:photonspectra}
\end{figure}

A sample of the simulated spectra are plotted in Fig.~\ref{fig:photonspectra}, where in all simulations the scattering pulse was assumed to be a plane wave with \textsc{fwhm} duration 45~fs and wavelength 800~nm.

Because we collide the electron and laser beams very close to the exit of the gas jet, the electron beam is significantly smaller than the laser beam.
Assuming that the electron beam waist is 1\,\si{\um} (based on measurements on Astra-Gemini and comparable to published results~\cite{Schnell2012}) and that the waist is positioned close to the edge of the gas jet (which is expected due to the focusing forces acting on the beam while it is inside the plasma), we calculate that the electron beam doubles in area over a distance of approximately 300\,\si{\um}.
At the collision point the electron beam area is therefore still approximately 1\,\si{\um}$^2$, compared to a laser beam area of 20\,\si{\um}$^2$.

 {Because the collision occurs between an electron bunch with finite duration and a laser pulse that is going through focus, different longitudinal slices of the electron bunch actually interact with different peak intensities.  
However, assuming a bunch length of $\approx \lambda_{\rm p}/2 \approx 10\,\si{\um}$ and using the measured variation of the laser intensity with distance from focus, we calculate that the difference in the peak $a_{\rm 0}$ experienced by the front and back of the bunch is less than 10\%.
The critical energy of the radiation spectrum varies as $\approx a_{\rm 0}^{0.5}$ so the effect of this on both the overall energy loss and radiation spectrum are small compared to the other fluctuations in the measurement.

The small transverse electron beam size and relatively slow longitudinal variation of $a_{\rm 0}$ along the electron bunch mean that radiation reaction is well described by a plane wave model, where one can neglect the variation in laser intensity due to focussing across the electron beam~\cite{Harvey2016}.}

\begin{figure}
	\includegraphics[width=\columnwidth]{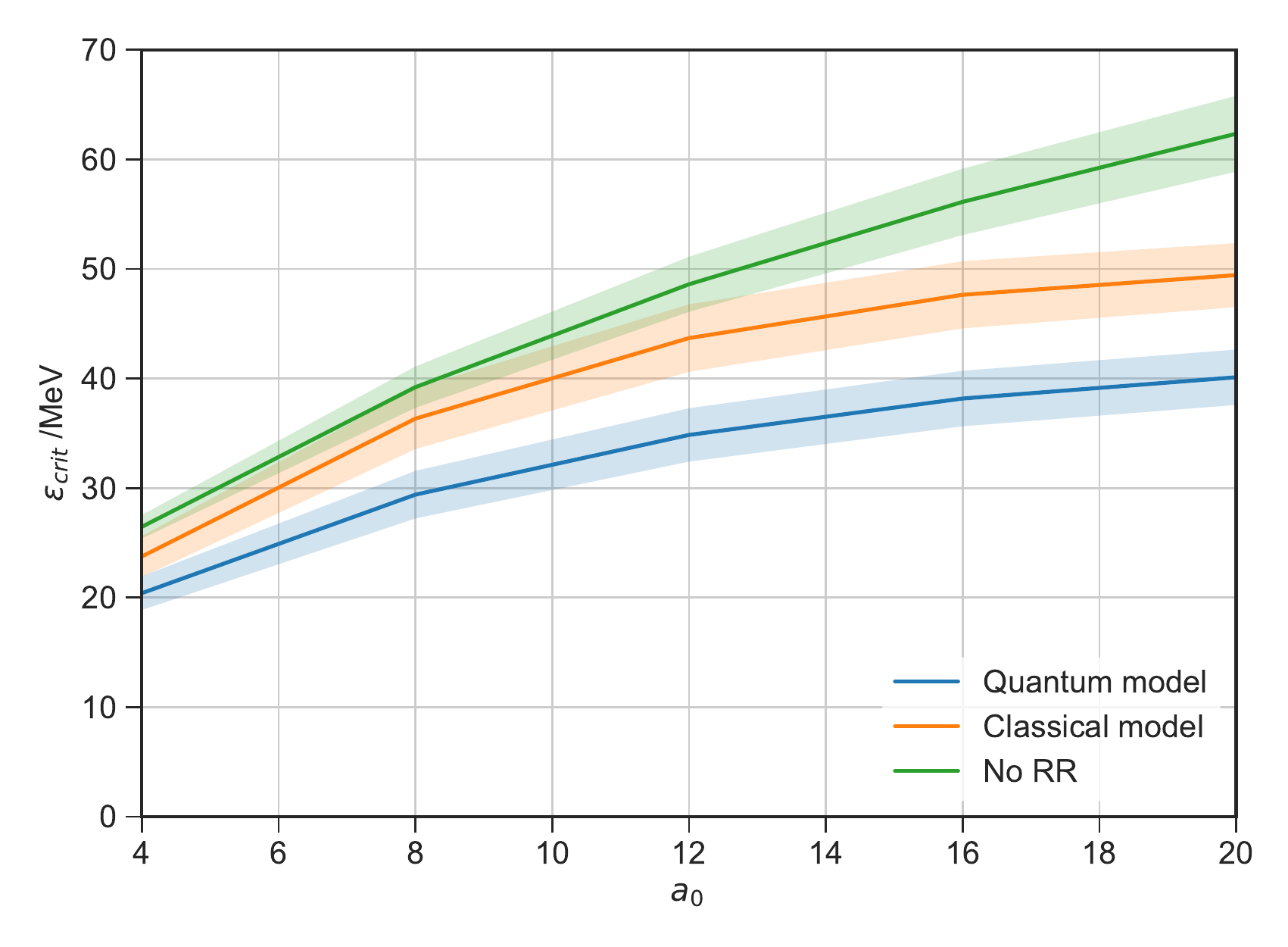}
	\caption{
		Simulated retrievals of $\varepsilon_\text{crit}$ assuming the collision of a plane wave of given peak $a_0$ with the experimentally measured electron spectra.
		The shaded regions represent $\pm 1 \sigma$ variations arising from the measured electron beam spectral fluctuations.
	}
	\label{fig:ecritvsa0}
\end{figure}

To assess the discriminatory power of our experiment against the different models, we simulated the full photon generation, measurement and fitting process for a range of peak $a_0$.
The $\varepsilon_\text{crit}$ which would be measured with a perfect noise-free detector are plotted in Fig.~\ref{fig:ecritvsa0}, where the results of the ICS simulations were interpolated over our measured `beam-off' electron spectra for the RR models, and the `beam-on' electron spectra for the `No RR' model.
For the `No RR' model the retrieved $\varepsilon_\text{crit}$ varies almost linearly with $a_0$, the models including radiation reaction predict a lower $\varepsilon_\text{crit}$  at high $a_0$ as expected.

%%%%%%%%%%%%%%%%%%%%%%%%%%%%%%%%%%%%%%%%%%%%%%
\section{Model comparison}
%%%%%%%%%%%%%%%%%%%%%%%%%%%%%%%%%%%%%%%%%%%%%%

\begin{figure}
	\includegraphics[width=\columnwidth]{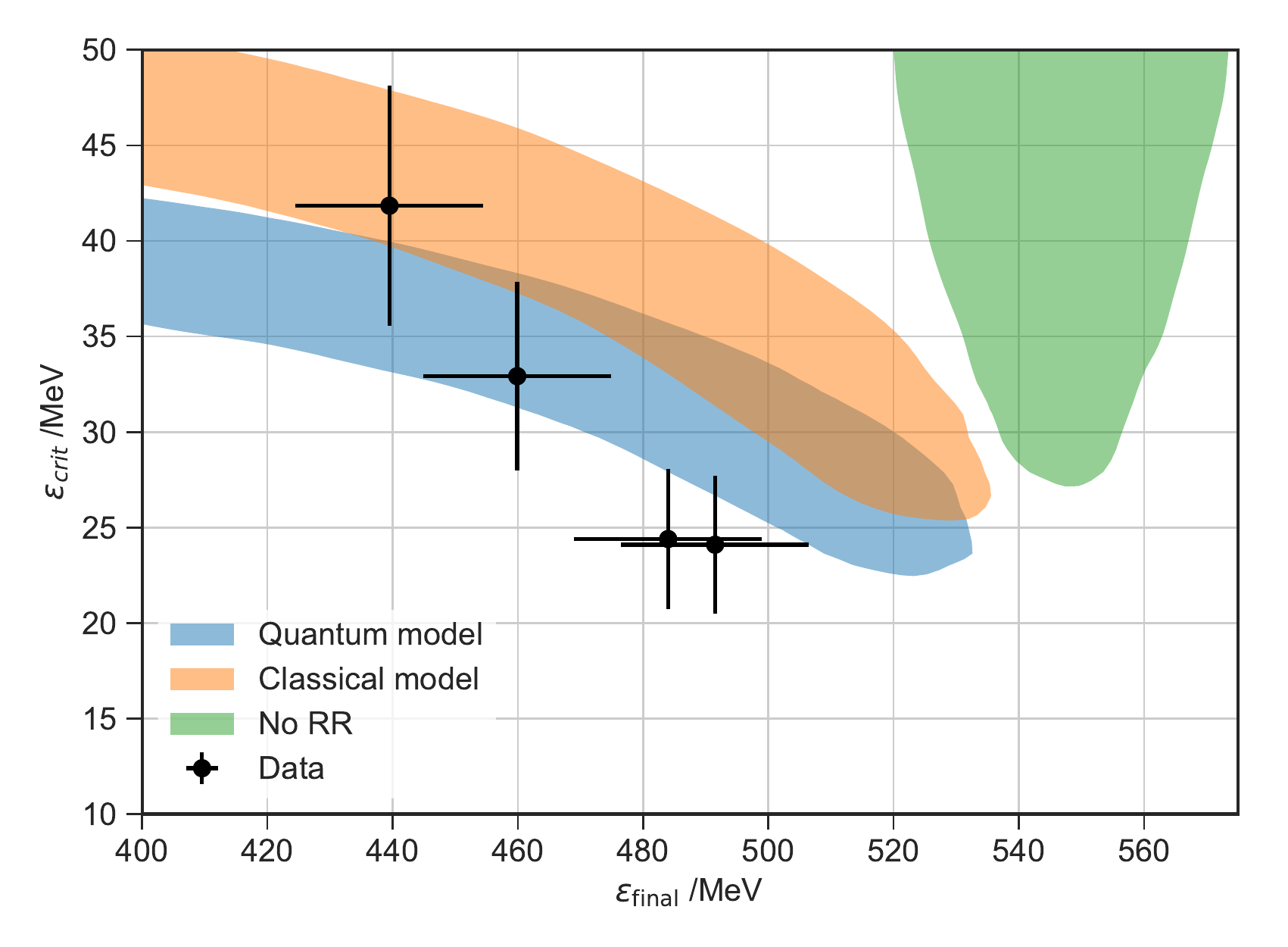}
	\caption{
			Experimentally measured $\varepsilon_\text{crit}$ as a function of $\varepsilon_{\rm final}$ measured at the electron spectral feature (points). 
			The shaded areas correspond to the results a hypothetical ensemble of identical experiments would measure 68\% of the time under different assumed radiation reaction models for a uniform distribution of $a_0$ between 4 and 20.
		}
	\label{fig:ecritvsef}
\end{figure}

	The measured $\varepsilon_\text{crit}$ is plotted as a function of the measured post-collision energy of the electron beam $\varepsilon_{\rm final}$ in Fig.~\ref{fig:ecritvsef} for the four shots, where the error bars represent diagnostic uncertainties.
	The sign of the correlation is significant, in that if we were observing Compton scattering without radiation reaction then the correlation should be in the positive direction.  
	After taking into account the measurement uncertainties in the data, we find that there is a $>$~98\% probability that the correlation is negative, the direction expected if radiation reaction is occurring in the successful collisions.
	Under the null hypothesis that no radiation reaction is occurring, the chance of observing a negative correlation at least this strong simultaneously with the electron energies all being below 500~MeV is lower than 1 in 3,000.

An important source of variation here is the interaction $a_0$, which should be expected to vary significantly between laser shots due primarily to the spatial jitter between the electron and photon beams.
If more data were available, one would therefore expect the points to trace out a curve in $(\Delta \varepsilon/\varepsilon, \varepsilon_\text{crit})$ space, parameterised along its length by $a_0$.
Each radiation reaction model generates a different curve, and so matching the data to a curve is a method for finding the model most consistent with the experiment independently of any knowledge of $a_0$ for a particular datum.

Due to the shot-to-shot jitter these curves are broadened into the shaded areas plotted in Fig.~\ref{fig:ecritvsef}.
To calculate these areas a large number of `numerical experiments' are conducted.
For various values of the laser strength $a_0$ (uniformly distributed over the range $a_0 = 4 - 20$), a set of 10 initial electron beam energies, $\varepsilon_{\rm initial}$ are drawn from the measured pre-collision distribution.
From the assumed $a_0$ a final energy after the interaction with the laser is calculated, $\varepsilon_{\rm final}$, and a set of $\varepsilon_{\rm crit}$ for each radiation reaction model as would be measured on the detector. 
The spread of of $\varepsilon_{\rm crit}$ takes into account the measured fluctuations of the electron beam spectral shape. 
We then calculate the averages $\overline{\varepsilon}_{\rm final}$ and $\overline{\varepsilon}_{\rm crit}$ and place these onto the ($\varepsilon_{\rm final}$, $\varepsilon_{\rm crit}$) space.  
This process is repeated 500 times for each value of $a_0$, equivalent to taking 50,000 successful collision shots.  
The shaded regions represent the area in ($\varepsilon_{\rm final}$, $\varepsilon_{\rm crit}$) space that contains 68\% of these numerical experiment results, i.e. what would be measured 68\% of the time under different radiation reaction models if one could repeat the experiment many times.

As $a_0$ tends to zero the $\gamma$-ray spectrum would become monochromatic.
Our $\gamma$-ray diagnostic would erroneously measure an finite effective $\varepsilon_{\rm crit}$ in this case, and for this reason the curves in Fig.~\ref{fig:ecritvsef} do not tend towards $\varepsilon_{\rm crit}=0$ at low $a_0$.

We observe that the data are more consistent with a quantum rather than classical model of radiation reaction, though there is large overlap between models at low $a_0$, and several data points are consistent with both models.
If the electron energy loss is ignored, it could be argued that the data are consistent with the `No RR' model if the interaction $a_0$ is lowered to $\sim$ 5.
However this situation is unlikely given the experimental precision of the spatial and temporal alignment between the electron bunch and colliding laser pulse and the observed correlation between electron beam energy, $\gamma$-ray yield and $\varepsilon_{\rm crit}$. 

As was discussed for the electron spectral data, it is also possible to estimate the interaction $a_0$ independently using the $\gamma$-ray spectra by interpolating the measured $\varepsilon_{\rm crit}$ onto the curves in Fig.~\ref{fig:ecritvsa0}.
We perform this estimation for each data point, and calculate the ratio of the estimates from the $\gamma$-ray data and the electron beam data $R = a_0(\varepsilon_\text{crit})/a_0(\varepsilon_{\rm final})$.
This is another metric of the model consistency which is independent of any knowledge of the interaction $a_0$.
The data is considered fixed so $R$ is a function of the model used to interpret the data, and perfect internal consistency implies $R = 1$.
Averaged over this data, at the 68\% level for the quantum model $R = 0.8^{+0.7}_{-0.3}$ and for the classical model $R=0.6^{+0.3}_{-0.2}$.
Under this metric the quantum radiation reaction model is slightly better at bringing the data from both diagnostics into agreement, whereas the classical model appears to systematically under-estimate $a_0$ for the $\gamma$-ray data compared to the electron beam data.

%%%%%%%%%%%%%%%%%%%%%%%%%%%%%%%%%%%%%%%%%%%%%%
\section{Discussion}
%%%%%%%%%%%%%%%%%%%%%%%%%%%%%%%%%%%%%%%%%%%%%%

The main evidence for the observation of radiation reaction presented here is the observation of low energy electron beams on all successful collision shots and the correlation between the post-collision electron beam energy and the $\gamma$-ray yield and spectrum.
This observation is consistent with the measurement of hard photons, of characteristic energy $\varepsilon_{\rm crit} > 30$~MeV, which carried a significant fraction of the initial electron energy meaning that the electron recoil should be non-negligible.
Moreover, this is reinforced by the agreement between the interaction $a_0$ inferred separately from the electron and $\gamma$-ray spectra under a quantum radiation reaction model, and that expected experimentally.

Simulations of the electron-laser overlap indicate that bright $\gamma$-ray beams with $\varepsilon_\text{crit} > 20$~MeV would be expected to be produced on 30\% of shots.
This is in line with our data when the measurements were taken immediately after alignment (4 out of 8 shots), though subsequent spatial and temporal drifts mean that the chance of later shots showing significant $\gamma$ signal drops quickly with time, limiting the duration of useful shot runs.
In future experiments we plan to more carefully identify and control these drifts, which will aid in the acquisition of a significantly larger data set.

While the observed correlation is encouraging, the number of shots demonstrating significant overlap in the experiment was limited and there are aspects of the laser-particle interaction which could be further investigated by the acquisition of more data. 
In classical radiation reaction the width of the electron energy spectrum can only decrease, but in quantum radiation reaction it can increase or decrease depending on the strength and duration of the interaction~\cite{Neitz2013,Vranic2016a,Ridgers2017}. 
This was not distinguishable here due to the  {value of $\eta$ achieved at collision}, the low number of shots and the variability in the electron spectra. 
 {Direct evidence of the stochastic nature of the photon emission during radiation reaction could be achieved in a similar  experimental set-up by increasing the electron beam stability and energy as well as the laser intensity at collision (e.g to $\approx 1.5$~GeV and $a_{\rm 0} \approx 15$).
}

We have focussed our attention on the energy loss of the dominant low-energy feature of the electron spectrum. 
The high-energy tail did not exhibit significant changes in energy, though removing it entirely from simulations of the inverse Compton scattering process generates photon beams with slightly lower $\varepsilon_\text{crit}$ indicating that it does in part contribute to the $\gamma$-ray beam despite its low charge.
Given that the two components of the electron spectra are so different, it is conceivable that each arises from a separate injection event inside the accelerator.
In this case it is plausible that the spectral components are separated in space and time inside and subsequently outside the plasma.
If so, it is possible that only a portion of the high-energy component experienced a significant interaction, which would diminish radiation reaction effects on that component of the beam.

%%%%%%%%%%%%%%%%%%%%%%%%%%%%%%%%%%%%%%%%%%%%%%
\section{Conclusion}
%%%%%%%%%%%%%%%%%%%%%%%%%%%%%%%%%%%%%%%%%%%%%%

In summary, we have presented data from a recent laser wakefield inverse Compton scattering experiment designed to identify the onset of radiation reaction. 
The electron and $\gamma$-ray spectra were simultaneously measured and independently used to infer the laser intensity during the interaction.
A fully quantum model of radiation reaction performs best in bringing the measurements from these two diagnostics into alignment.
Furthermore, we have generated $\gamma$-ray beams with the highest energies yet reported from an all-optical inverse Compton scattering scheme, previously limited to below 20~MeV~\cite{Sarri2014, Yan2017}, and measurable here with a scintillation detector highly sensitive to the electromagnetic shower produced by high energy photons.

 {While the results presented here represent statistically significant evidence of radiation reaction occurring during the collision of a high-intensity laser pulse with a high-energy electron beam, they are not a systematic study of radiation reaction.  
The QED model currently appears to provide a more self-consistent description of the data than a purely classical one, however this is only at the 1$\sigma$ level.  
A future study systematically varying the quantum parameter $\eta$ through either the electron beam energy or laser intensity which extends to higher values of $\eta$ than achieved in this experiment will allow detailed comparisons of different models and a proper assessment of their range of validity. 
Laser wakefield accelerators have already demonstrated sufficient beam stability, e.g. \cite{Faure2006,Buck2013, Gonsalves2011}, and higher beam energies, e.g. \cite{Leemans2014}, and higher laser intensities are readily achievable with existing lasers, so that the necessary $\eta$ to clearly observe quantum radiation reaction is within reach of existing laser systems.
The main remaining challenge is therefore to achieve all of these simultaneously in an experiment which maintains femtosecond and micrometre level alignment over extended periods of time, so as to allow sufficient data to be collected for a systematic study of radiation reaction in the quantum regime.
%The measurements presented here demonstrate that, with improved control of alignment drifts and more stable electron injection \cite{something appropriate here}, such a study will be readily achievable on existing laser systems.
}

%%%%%%%%%%%%%%%%%%%%%%%%%%%%%%%%%%%%%%%%%%%%%%
\section{Data availability}
%%%%%%%%%%%%%%%%%%%%%%%%%%%%%%%%%%%%%%%%%%%%%% 

The authors confirm that the all data used in this study are available without restriction. Data can be obtained by contacting plasma@imperial.ac.uk.

%%%%%%%%%%%%%%%%%%%%%%%%%%%%%%%%%%%%%%%%%%%%%%
\section{Acknowledgements}
%%%%%%%%%%%%%%%%%%%%%%%%%%%%%%%%%%%%%%%%%%%%%%

We acknowledge funding from EPSRC grants EP/M018555/1, EP/M018091/1 and EP/M018156/1, STFC grants ST/J002062/1 and ST/P000835/1, Horizon 2020 funding under the Marie Sk\l odowska-Curie grant No.~701676 (A.I.) and the European Research Council (ERC) grant agreement No. 682399 (S.M.), the Knut \& Alice Wallenberg Foundation (T.B., C.H., A.I, M.M.), the Swedish Research Council, grants 2012-5644 and 2013-4248 (M.M.), the US NSF CAREER Award 1054164 (A.T., K.B.), and US DOD under grant W911NF1610044 (A.T., K.B., A.J., K.K.). Simulations were performed on resources provided by the Swedish National Infrastructure for Computing at the HPC2N. We would like to thank the CLF for their assistance in running the experiment.

\bibliography{Mendeley_ICS_Paper_10authors}

%merlin.mbs 2010-03-15 4.21a (PWD, AO, DPC)
\begin{thebibliography}{57}%
\makeatletter
\providecommand \@ifxundefined [1]{%
 \@ifx{#1\undefined}
}%
\providecommand \@ifnum [1]{%
 \ifnum #1\expandafter \@firstoftwo
 \else \expandafter \@secondoftwo
 \fi
}%
\providecommand \@ifx [1]{%
 \ifx #1\expandafter \@firstoftwo
 \else \expandafter \@secondoftwo
 \fi
}%
\providecommand \natexlab [1]{#1}%
\providecommand \enquote  [1]{``#1''}%
\providecommand \bibnamefont  [1]{#1}%
\providecommand \bibfnamefont [1]{#1}%
\providecommand \citenamefont [1]{#1}%
\providecommand \href@noop [0]{\@secondoftwo}%
\providecommand \href [0]{\begingroup \@sanitize@url \@href}%
\providecommand \@href[1]{\@@startlink{#1}\@@href}%
\providecommand \@@href[1]{\endgroup#1\@@endlink}%
\providecommand \@sanitize@url [0]{\catcode `\\12\catcode `\$12\catcode
  `\&12\catcode `\#12\catcode `\^12\catcode `\_12\catcode `\%12\relax}%
\providecommand \@@startlink[1]{}%
\providecommand \@@endlink[0]{}%
\providecommand \url  [0]{\begingroup\@sanitize@url \@url }%
\providecommand \@url [1]{\endgroup\@href {#1}{\urlprefix }}%
\providecommand \urlprefix  [0]{URL }%
\providecommand \Eprint [0]{\href }%
\@ifxundefined \urlstyle {%
  \providecommand \doi  [0]{\begingroup \@sanitize@url \@doi}%
  \providecommand \@doi [1]{\endgroup \@@startlink {\doibase
  #1}doi:\discretionary {}{}{}#1\@@endlink }%
}{%
  \providecommand \doi  [0]{doi:\discretionary{}{}{}\begingroup
  \urlstyle{rm}\Url }%
}%
\providecommand \doibase [0]{http://dx.doi.org/}%
\providecommand \Doi [0]{\begingroup \@sanitize@url \@Doi }%
\providecommand \@Doi  [1]{\endgroup\@@startlink{\doibase#1}\@@Doi}%
\providecommand \@@Doi [1]{#1\@@endlink}%
\providecommand \selectlanguage [0]{\@gobble}%
\providecommand \bibinfo  [0]{\@secondoftwo}%
\providecommand \bibfield  [0]{\@secondoftwo}%
\providecommand \translation [1]{[#1]}%
\providecommand \BibitemOpen [0]{}%
\providecommand \bibitemStop [0]{}%
\providecommand \bibitemNoStop [0]{.\EOS\space}%
\providecommand \EOS [0]{\spacefactor3000\relax}%
\providecommand \BibitemShut  [1]{\csname bibitem#1\endcsname}%
%</preamble>
\bibitem [{\citenamefont {Tavani}\ \emph {et~al.}(2011)\citenamefont {Tavani},
  \citenamefont {Bulgarelli}, \citenamefont {Vittorini}, \citenamefont
  {Pellizzoni}, \citenamefont {Striani}, \citenamefont {Caraveo}, \citenamefont
  {Weisskopf}, \citenamefont {Tennant}, \citenamefont {Pucella}, \citenamefont
  {Trois} \emph {et~al.}}]{Tchufistova2011}%
  \BibitemOpen
  \bibfield  {author} {\bibinfo {author} {\bibfnamefont {M.}~\bibnamefont
  {Tavani}}, \bibinfo {author} {\bibfnamefont {A.}~\bibnamefont {Bulgarelli}},
  \bibinfo {author} {\bibfnamefont {V.}~\bibnamefont {Vittorini}}, \bibinfo
  {author} {\bibfnamefont {A.}~\bibnamefont {Pellizzoni}}, \bibinfo {author}
  {\bibfnamefont {E.}~\bibnamefont {Striani}}, \bibinfo {author} {\bibfnamefont
  {P.}~\bibnamefont {Caraveo}}, \bibinfo {author} {\bibfnamefont {M.~C.}\
  \bibnamefont {Weisskopf}}, \bibinfo {author} {\bibfnamefont {A.}~\bibnamefont
  {Tennant}}, \bibinfo {author} {\bibfnamefont {G.}~\bibnamefont {Pucella}},
  \bibinfo {author} {\bibfnamefont {A.}~\bibnamefont {Trois}},  \emph
  {et~al.},\ }\bibfield  {title} {\enquote {\bibinfo {title} {{Discovery of
  Powerful Gamma-Ray Flares from the Crab Nebula}},}\ }\Doi
  {10.1126/science.1200083} {\bibfield  {journal} {\bibinfo  {journal}
  {Science},\ }\textbf {\bibinfo {volume} {331}},\ \bibinfo {pages} {736}
  (\bibinfo {year} {2011})}\BibitemShut {NoStop}%
\bibitem [{\citenamefont {Jaroschek}\ and\ \citenamefont
  {Hoshino}(2009)}]{Jaroschek2009}%
  \BibitemOpen
  \bibfield  {author} {\bibinfo {author} {\bibfnamefont {C.~H.}\ \bibnamefont
  {Jaroschek}}\ and\ \bibinfo {author} {\bibfnamefont {M.}~\bibnamefont
  {Hoshino}},\ }\bibfield  {title} {\enquote {\bibinfo {title}
  {{Radiation-Dominated Relativistic Current Sheets}},}\ }\Doi
  {10.1103/PhysRevLett.103.075002} {\bibfield  {journal} {\bibinfo  {journal}
  {Phys. Rev. Lett.},\ }\textbf {\bibinfo {volume} {103}},\ \bibinfo {pages}
  {075002} (\bibinfo {year} {2009})}\BibitemShut {NoStop}%
\bibitem [{\citenamefont {Di~Piazza}\ \emph {et~al.}(2012)\citenamefont
  {Di~Piazza}, \citenamefont {Muller}, \citenamefont {Hatsagortsyan},\ and\
  \citenamefont {Keitel}}]{DiPiazza2012a}%
  \BibitemOpen
  \bibfield  {author} {\bibinfo {author} {\bibfnamefont {A.}~\bibnamefont
  {Di~Piazza}}, \bibinfo {author} {\bibfnamefont {C.}~\bibnamefont {Muller}},
  \bibinfo {author} {\bibfnamefont {K.~Z.}\ \bibnamefont {Hatsagortsyan}}, \
  and\ \bibinfo {author} {\bibfnamefont {C.~H.}\ \bibnamefont {Keitel}},\
  }\bibfield  {title} {\enquote {\bibinfo {title} {{Extremely high-intensity
  laser interactions with fundamental quantum systems}},}\ }\Doi
  {10.1103/RevModPhys.84.1177} {\bibfield  {journal} {\bibinfo  {journal} {Rev.
  Mod. Phys.},\ }\textbf {\bibinfo {volume} {84}},\ \bibinfo {pages} {1177}
  (\bibinfo {year} {2012})}\BibitemShut {NoStop}%
\bibitem [{\citenamefont {Thomas}\ \emph {et~al.}(2012)\citenamefont {Thomas},
  \citenamefont {Ridgers}, \citenamefont {Bulanov}, \citenamefont {Griffin},\
  and\ \citenamefont {Mangles}}]{Thomas2012}%
  \BibitemOpen
  \bibfield  {author} {\bibinfo {author} {\bibfnamefont {A.~G.~R.}\
  \bibnamefont {Thomas}}, \bibinfo {author} {\bibfnamefont {C.~P.}\
  \bibnamefont {Ridgers}}, \bibinfo {author} {\bibfnamefont {S.~S.}\
  \bibnamefont {Bulanov}}, \bibinfo {author} {\bibfnamefont {B.~J.}\
  \bibnamefont {Griffin}}, \ and\ \bibinfo {author} {\bibfnamefont {S.~P.~D.}\
  \bibnamefont {Mangles}},\ }\bibfield  {title} {\enquote {\bibinfo {title}
  {{Strong Radiation-Damping Effects in a Gamma-Ray Source Generated by the
  Interaction of a High-Intensity Laser with a Wakefield-Accelerated Electron
  Beam}},}\ }\Doi {10.1103/PhysRevX.2.041004} {\bibfield  {journal} {\bibinfo
  {journal} {Phys. Rev. X},\ }\textbf {\bibinfo {volume} {2}},\ \bibinfo
  {pages} {041004} (\bibinfo {year} {2012})}\BibitemShut {NoStop}%
\bibitem [{\citenamefont {Hammond}(2010)}]{Hammond2010}%
  \BibitemOpen
  \bibfield  {author} {\bibinfo {author} {\bibfnamefont {R.}~\bibnamefont
  {Hammond}},\ }\bibfield  {title} {\enquote {\bibinfo {title} {{Radiation
  reaction at ultrahigh intensities}},}\ }\Doi {10.1103/PhysRevA.81.062104}
  {\bibfield  {journal} {\bibinfo  {journal} {Phys. Rev. A},\ }\textbf
  {\bibinfo {volume} {81}},\ \bibinfo {pages} {062104} (\bibinfo {year}
  {2010})}\BibitemShut {NoStop}%
\bibitem [{\citenamefont {Burton}\ and\ \citenamefont
  {Noble}(2014)}]{Burton2014b}%
  \BibitemOpen
  \bibfield  {author} {\bibinfo {author} {\bibfnamefont {D.~A.}\ \bibnamefont
  {Burton}}\ and\ \bibinfo {author} {\bibfnamefont {A.}~\bibnamefont {Noble}},\
  }\bibfield  {title} {\enquote {\bibinfo {title} {{Aspects of electromagnetic
  radiation reaction in strong fields}},}\ }\Doi {10.1080/00107514.2014.886840}
  {\bibfield  {journal} {\bibinfo  {journal} {Contemp. Phys.},\ }\textbf
  {\bibinfo {volume} {55}},\ \bibinfo {pages} {110} (\bibinfo {year}
  {2014})}\BibitemShut {NoStop}%
\bibitem [{\citenamefont {Landau}\ and\ \citenamefont
  {Lifshitz}(1975)}]{Landau1975}%
  \BibitemOpen
  \bibfield  {author} {\bibinfo {author} {\bibfnamefont {L.~D.}\ \bibnamefont
  {Landau}}\ and\ \bibinfo {author} {\bibfnamefont {E.~M.}\ \bibnamefont
  {Lifshitz}},\ }\href@noop {} {\emph {\bibinfo {title} {{The Classical Theory
  of Fields}}}}\ (\bibinfo  {publisher} {Elsevier},\ \bibinfo {address}
  {Oxford},\ \bibinfo {year} {1975})\BibitemShut {NoStop}%
\bibitem [{\citenamefont {KrivitskiÄ­}\ and\ \citenamefont
  {Tsytovich}(1991)}]{Krivitskii1991}%
  \BibitemOpen
  \bibfield  {author} {\bibinfo {author} {\bibfnamefont {V.~S.}\ \bibnamefont
  {KrivitskiÄ­}}\ and\ \bibinfo {author} {\bibfnamefont {V.~N.}\ \bibnamefont
  {Tsytovich}},\ }\bibfield  {title} {\enquote {\bibinfo {title} {{Average
  radiation-reaction force in quantum electrodynamics}},}\ }\Doi
  {10.1070/PU1991v034n03ABEH002352} {\bibfield  {journal} {\bibinfo  {journal}
  {Sov. Phys. Usp.},\ }\textbf {\bibinfo {volume} {34}},\ \bibinfo {pages}
  {250} (\bibinfo {year} {1991})}\BibitemShut {NoStop}%
\bibitem [{\citenamefont {Ilderton}\ and\ \citenamefont
  {Torgrimsson}(2013)}]{Ilderton2013}%
  \BibitemOpen
  \bibfield  {author} {\bibinfo {author} {\bibfnamefont {A.}~\bibnamefont
  {Ilderton}}\ and\ \bibinfo {author} {\bibfnamefont {G.}~\bibnamefont
  {Torgrimsson}},\ }\bibfield  {title} {\enquote {\bibinfo {title} {{Radiation
  reaction from QED: Lightfront perturbation theory in a plane wave
  background}},}\ }\Doi {10.1103/PhysRevD.88.025021} {\bibfield  {journal}
  {\bibinfo  {journal} {Phys. Rev. D},\ }\textbf {\bibinfo {volume} {88}},\
  \bibinfo {pages} {025021} (\bibinfo {year} {2013})}\BibitemShut {NoStop}%
\bibitem [{\citenamefont {Neitz}\ and\ \citenamefont
  {Di~Piazza}(2013)}]{Neitz2013}%
  \BibitemOpen
  \bibfield  {author} {\bibinfo {author} {\bibfnamefont {N.}~\bibnamefont
  {Neitz}}\ and\ \bibinfo {author} {\bibfnamefont {A.}~\bibnamefont
  {Di~Piazza}},\ }\bibfield  {title} {\enquote {\bibinfo {title}
  {{Stochasticity effects in quantum radiation reaction}},}\ }\Doi
  {10.1103/PhysRevLett.111.054802} {\bibfield  {journal} {\bibinfo  {journal}
  {Phys. Rev. Lett.},\ }\textbf {\bibinfo {volume} {111}},\ \bibinfo {pages}
  {054802} (\bibinfo {year} {2013})}\BibitemShut {NoStop}%
\bibitem [{\citenamefont {Yoffe}\ \emph {et~al.}(2015)\citenamefont {Yoffe},
  \citenamefont {Kravets}, \citenamefont {Noble},\ and\ \citenamefont
  {Jaroszynski}}]{Yoffe2015}%
  \BibitemOpen
  \bibfield  {author} {\bibinfo {author} {\bibfnamefont {S.~R.}\ \bibnamefont
  {Yoffe}}, \bibinfo {author} {\bibfnamefont {Y.}~\bibnamefont {Kravets}},
  \bibinfo {author} {\bibfnamefont {A.}~\bibnamefont {Noble}}, \ and\ \bibinfo
  {author} {\bibfnamefont {D.~A.}\ \bibnamefont {Jaroszynski}},\ }\bibfield
  {title} {\enquote {\bibinfo {title} {{Longitudinal and transverse cooling of
  relativistic electron beams in intense laser pulses}},}\ }\Doi
  {10.1088/1367-2630/17/5/053025} {\bibfield  {journal} {\bibinfo  {journal}
  {New. J. Phys.},\ }\textbf {\bibinfo {volume} {17}},\ \bibinfo {pages}
  {053025} (\bibinfo {year} {2015})}\BibitemShut {NoStop}%
\bibitem [{\citenamefont {Vranic}\ \emph
  {et~al.}(2016){\natexlab{a}}\citenamefont {Vranic}, \citenamefont
  {Grismayer}, \citenamefont {Fonseca},\ and\ \citenamefont
  {Silva}}]{Vranic2016a}%
  \BibitemOpen
  \bibfield  {author} {\bibinfo {author} {\bibfnamefont {M.}~\bibnamefont
  {Vranic}}, \bibinfo {author} {\bibfnamefont {T.}~\bibnamefont {Grismayer}},
  \bibinfo {author} {\bibfnamefont {R.~A.}\ \bibnamefont {Fonseca}}, \ and\
  \bibinfo {author} {\bibfnamefont {L.~O.}\ \bibnamefont {Silva}},\ }\bibfield
  {title} {\enquote {\bibinfo {title} {{Quantum radiation reaction in head-on
  laser-electron beam interaction}},}\ }\Doi {10.1088/1367-2630/18/7/073035}
  {\bibfield  {journal} {\bibinfo  {journal} {New. J. Phys.},\ }\textbf
  {\bibinfo {volume} {18}},\ \bibinfo {pages} {073035} (\bibinfo {year}
  {2016}{\natexlab{a}})}\BibitemShut {NoStop}%
\bibitem [{\citenamefont {Dinu}\ \emph {et~al.}(2016)\citenamefont {Dinu},
  \citenamefont {Harvey}, \citenamefont {Ilderton}, \citenamefont {Marklund},\
  and\ \citenamefont {Torgrimsson}}]{Dinu2016}%
  \BibitemOpen
  \bibfield  {author} {\bibinfo {author} {\bibfnamefont {V.}~\bibnamefont
  {Dinu}}, \bibinfo {author} {\bibfnamefont {C.}~\bibnamefont {Harvey}},
  \bibinfo {author} {\bibfnamefont {A.}~\bibnamefont {Ilderton}}, \bibinfo
  {author} {\bibfnamefont {M.}~\bibnamefont {Marklund}}, \ and\ \bibinfo
  {author} {\bibfnamefont {G.}~\bibnamefont {Torgrimsson}},\ }\bibfield
  {title} {\enquote {\bibinfo {title} {{Quantum Radiation Reaction: From
  Interference to Incoherence}},}\ }\Doi {10.1103/PhysRevLett.116.044801}
  {\bibfield  {journal} {\bibinfo  {journal} {Phys. Rev. Lett.},\ }\textbf
  {\bibinfo {volume} {116}},\ \bibinfo {pages} {044801} (\bibinfo {year}
  {2016})}\BibitemShut {NoStop}%
\bibitem [{\citenamefont {Mangles}\ \emph {et~al.}(2004)\citenamefont
  {Mangles}, \citenamefont {Murphy}, \citenamefont {Najmudin}, \citenamefont
  {Thomas}, \citenamefont {Collier}, \citenamefont {Dangor}, \citenamefont
  {Divall}, \citenamefont {Foster}, \citenamefont {Gallacher}, \citenamefont
  {Hooker} \emph {et~al.}}]{Mangles2004a}%
  \BibitemOpen
  \bibfield  {author} {\bibinfo {author} {\bibfnamefont {S.~P.~D.}\
  \bibnamefont {Mangles}}, \bibinfo {author} {\bibfnamefont {C.~D.}\
  \bibnamefont {Murphy}}, \bibinfo {author} {\bibfnamefont {Z.}~\bibnamefont
  {Najmudin}}, \bibinfo {author} {\bibfnamefont {A.~G.~R.}\ \bibnamefont
  {Thomas}}, \bibinfo {author} {\bibfnamefont {J.~L.}\ \bibnamefont {Collier}},
  \bibinfo {author} {\bibfnamefont {A.~E.}\ \bibnamefont {Dangor}}, \bibinfo
  {author} {\bibfnamefont {E.~J.}\ \bibnamefont {Divall}}, \bibinfo {author}
  {\bibfnamefont {P.~S.}\ \bibnamefont {Foster}}, \bibinfo {author}
  {\bibfnamefont {J.~G.}\ \bibnamefont {Gallacher}}, \bibinfo {author}
  {\bibfnamefont {C.~J.}\ \bibnamefont {Hooker}},  \emph {et~al.},\ }\bibfield
  {title} {\enquote {\bibinfo {title} {{Monoenergetic beams of relativistic
  electrons from intense laser-plasma interactions.}}}\ }\href
  {http://www.nature.com/nature/journal/v431/n7008/abs/nature02939.html}
  {\bibfield  {journal} {\bibinfo  {journal} {Nature},\ }\textbf {\bibinfo
  {volume} {431}},\ \bibinfo {pages} {535} (\bibinfo {year}
  {2004})}\BibitemShut {NoStop}%
\bibitem [{\citenamefont {Geddes}\ \emph {et~al.}(2004)\citenamefont {Geddes},
  \citenamefont {Toth}, \citenamefont {Van~Tilborg}, \citenamefont {Esarey},
  \citenamefont {Schroeder}, \citenamefont {Bruhwiler}, \citenamefont {Nieter},
  \citenamefont {Cary},\ and\ \citenamefont {Leemans}}]{Geddes2004}%
  \BibitemOpen
  \bibfield  {author} {\bibinfo {author} {\bibfnamefont {C.~G.~R.}\
  \bibnamefont {Geddes}}, \bibinfo {author} {\bibfnamefont {C.~S.}\
  \bibnamefont {Toth}}, \bibinfo {author} {\bibfnamefont {J.}~\bibnamefont
  {Van~Tilborg}}, \bibinfo {author} {\bibfnamefont {E.}~\bibnamefont {Esarey}},
  \bibinfo {author} {\bibfnamefont {C.~B.}\ \bibnamefont {Schroeder}}, \bibinfo
  {author} {\bibfnamefont {D.}~\bibnamefont {Bruhwiler}}, \bibinfo {author}
  {\bibfnamefont {C.}~\bibnamefont {Nieter}}, \bibinfo {author} {\bibfnamefont
  {J.}~\bibnamefont {Cary}}, \ and\ \bibinfo {author} {\bibfnamefont {W.~P.}\
  \bibnamefont {Leemans}},\ }\bibfield  {title} {\enquote {\bibinfo {title}
  {{High-quality electron beams from a laser wakefield accelerator using
  plasma-channel guiding.}}}\ }\href
  {http://www.nature.com/nature/journal/v431/n7008/abs/nature02900.html}
  {\bibfield  {journal} {\bibinfo  {journal} {Nature},\ }\textbf {\bibinfo
  {volume} {431}},\ \bibinfo {pages} {538} (\bibinfo {year}
  {2004})}\BibitemShut {NoStop}%
\bibitem [{\citenamefont {Faure}\ \emph {et~al.}(2004)\citenamefont {Faure},
  \citenamefont {Glinec}, \citenamefont {Pukhov}, \citenamefont {Kiselev},
  \citenamefont {Gordienko}, \citenamefont {Lefebvre}, \citenamefont
  {Rousseau}, \citenamefont {Burgy},\ and\ \citenamefont {Malka}}]{Faure2004}%
  \BibitemOpen
  \bibfield  {author} {\bibinfo {author} {\bibfnamefont {J.}~\bibnamefont
  {Faure}}, \bibinfo {author} {\bibfnamefont {Y.}~\bibnamefont {Glinec}},
  \bibinfo {author} {\bibfnamefont {A.}~\bibnamefont {Pukhov}}, \bibinfo
  {author} {\bibfnamefont {S.}~\bibnamefont {Kiselev}}, \bibinfo {author}
  {\bibfnamefont {S.}~\bibnamefont {Gordienko}}, \bibinfo {author}
  {\bibfnamefont {E.}~\bibnamefont {Lefebvre}}, \bibinfo {author}
  {\bibfnamefont {J-P.}\ \bibnamefont {Rousseau}}, \bibinfo {author}
  {\bibfnamefont {F.}~\bibnamefont {Burgy}}, \ and\ \bibinfo {author}
  {\bibfnamefont {V.}~\bibnamefont {Malka}},\ }\bibfield  {title} {\enquote
  {\bibinfo {title} {{A laser-plasma accelerator producing monoenergetic
  electron beams.}}}\ }\href
  {http://www.nature.com/nature/journal/v431/n7008/abs/nature02963.html}
  {\bibfield  {journal} {\bibinfo  {journal} {Nature},\ }\textbf {\bibinfo
  {volume} {431}},\ \bibinfo {pages} {541} (\bibinfo {year}
  {2004})}\BibitemShut {NoStop}%
\bibitem [{\citenamefont {Esarey}\ \emph {et~al.}(2009)\citenamefont {Esarey},
  \citenamefont {Schroeder},\ and\ \citenamefont {Leemans}}]{Esarey2009a}%
  \BibitemOpen
  \bibfield  {author} {\bibinfo {author} {\bibfnamefont {E.}~\bibnamefont
  {Esarey}}, \bibinfo {author} {\bibfnamefont {C.~B.}\ \bibnamefont
  {Schroeder}}, \ and\ \bibinfo {author} {\bibfnamefont {W.~P.}\ \bibnamefont
  {Leemans}},\ }\bibfield  {title} {\enquote {\bibinfo {title} {{Physics of
  laser-driven plasma-based electron accelerators}},}\ }\Doi
  {10.1103/RevModPhys.81.1229} {\bibfield  {journal} {\bibinfo  {journal} {Rev.
  Mod. Phys.},\ }\textbf {\bibinfo {volume} {81}},\ \bibinfo {pages} {1229}
  (\bibinfo {year} {2009})}\BibitemShut {NoStop}%
\bibitem [{\citenamefont {Leemans}\ \emph {et~al.}(2006)\citenamefont
  {Leemans}, \citenamefont {Nagler}, \citenamefont {Gonsalves}, \citenamefont
  {Toth}, \citenamefont {Nakamura}, \citenamefont {Geddes}, \citenamefont
  {Esarey}, \citenamefont {Schroeder},\ and\ \citenamefont
  {Hooker}}]{Leemans2006}%
  \BibitemOpen
  \bibfield  {author} {\bibinfo {author} {\bibfnamefont {W.P.}\ \bibnamefont
  {Leemans}}, \bibinfo {author} {\bibfnamefont {B}~\bibnamefont {Nagler}},
  \bibinfo {author} {\bibfnamefont {A.~J.}\ \bibnamefont {Gonsalves}}, \bibinfo
  {author} {\bibfnamefont {C}~\bibnamefont {Toth}}, \bibinfo {author}
  {\bibfnamefont {K}~\bibnamefont {Nakamura}}, \bibinfo {author} {\bibfnamefont
  {C.~G.~R.}\ \bibnamefont {Geddes}}, \bibinfo {author} {\bibfnamefont
  {E}~\bibnamefont {Esarey}}, \bibinfo {author} {\bibfnamefont {C.~B.}\
  \bibnamefont {Schroeder}}, \ and\ \bibinfo {author} {\bibfnamefont {S.~M.}\
  \bibnamefont {Hooker}},\ }\bibfield  {title} {\enquote {\bibinfo {title}
  {{GeV electron beams from a centimetre-scale accelerator}},}\ }\href@noop {}
  {\bibfield  {journal} {\bibinfo  {journal} {Nature Physics},\ }\textbf
  {\bibinfo {volume} {2}},\ \bibinfo {pages} {696--699} (\bibinfo {year}
  {2006})}\BibitemShut {NoStop}%
\bibitem [{\citenamefont {Kneip}\ \emph {et~al.}(2009)\citenamefont {Kneip},
  \citenamefont {Nagel}, \citenamefont {Martins}, \citenamefont {Mangles},
  \citenamefont {Bellei}, \citenamefont {Chekhlov}, \citenamefont {Clarke},
  \citenamefont {Delerue}, \citenamefont {Divall}, \citenamefont {Doucas} \emph
  {et~al.}}]{Kneip2009}%
  \BibitemOpen
  \bibfield  {author} {\bibinfo {author} {\bibfnamefont {S.}~\bibnamefont
  {Kneip}}, \bibinfo {author} {\bibfnamefont {S.R.}\ \bibnamefont {Nagel}},
  \bibinfo {author} {\bibfnamefont {S.F.}\ \bibnamefont {Martins}}, \bibinfo
  {author} {\bibfnamefont {S. P. D.}\ \bibnamefont {Mangles}}, \bibinfo
  {author} {\bibfnamefont {C.}~\bibnamefont {Bellei}}, \bibinfo {author}
  {\bibfnamefont {O.}~\bibnamefont {Chekhlov}}, \bibinfo {author}
  {\bibfnamefont {R.J.}\ \bibnamefont {Clarke}}, \bibinfo {author}
  {\bibfnamefont {N.}~\bibnamefont {Delerue}}, \bibinfo {author} {\bibfnamefont
  {E.J.}\ \bibnamefont {Divall}}, \bibinfo {author} {\bibfnamefont
  {G.}~\bibnamefont {Doucas}},  \emph {et~al.},\ }\bibfield  {title} {\enquote
  {\bibinfo {title} {{Near-GeV acceleration of electrons by a nonlinear plasma
  wave driven by a self-guided laser pulse}},}\ }\Doi
  {10.1103/PhysRevLett.103.035002} {\bibfield  {journal} {\bibinfo  {journal}
  {Physical Review Letters},\ }\textbf {\bibinfo {volume} {103}} (\bibinfo
  {year} {2009})},\ \doi {10.1103/PhysRevLett.103.035002}\BibitemShut {NoStop}%
\bibitem [{\citenamefont {Leemans}\ \emph {et~al.}(2014)\citenamefont
  {Leemans}, \citenamefont {Gonsalves}, \citenamefont {Mao}, \citenamefont
  {Nakamura}, \citenamefont {Benedetti}, \citenamefont {Schroeder},
  \citenamefont {Tth}, \citenamefont {Daniels}, \citenamefont {Mittelberger},\
  and\ \citenamefont {Bulanov}}]{Leemans2014}%
  \BibitemOpen
  \bibfield  {author} {\bibinfo {author} {\bibfnamefont {W.P.}\ \bibnamefont
  {Leemans}}, \bibinfo {author} {\bibfnamefont {A.~J.}\ \bibnamefont
  {Gonsalves}}, \bibinfo {author} {\bibfnamefont {H.-S.}\ \bibnamefont {Mao}},
  \bibinfo {author} {\bibfnamefont {K}~\bibnamefont {Nakamura}}, \bibinfo
  {author} {\bibfnamefont {C}~\bibnamefont {Benedetti}}, \bibinfo {author}
  {\bibfnamefont {C.~B.}\ \bibnamefont {Schroeder}}, \bibinfo {author}
  {\bibfnamefont {Cs.}\ \bibnamefont {Tth}}, \bibinfo {author} {\bibfnamefont
  {J}~\bibnamefont {Daniels}}, \bibinfo {author} {\bibfnamefont {D~E}\
  \bibnamefont {Mittelberger}}, \ and\ \bibinfo {author} {\bibfnamefont {S~S}\
  \bibnamefont {Bulanov}},\ }\bibfield  {title} {\enquote {\bibinfo {title}
  {{Multi-GeV Electron Beams from Capillary-Discharge-Guided Subpetawatt Laser
  Pulses in the Self-Trapping Regime}},}\ }\Doi
  {10.1103/physrevlett.113.245002} {\bibfield  {journal} {\bibinfo  {journal}
  {Phys. Rev. Lett.},\ }\textbf {\bibinfo {volume} {245002}},\ \bibinfo {pages}
  {1--5} (\bibinfo {year} {2014})}\BibitemShut {NoStop}%
\bibitem [{\citenamefont {Wang}\ \emph {et~al.}(2013)\citenamefont {Wang},
  \citenamefont {Zgadzaj}, \citenamefont {Fazel}, \citenamefont {Li},
  \citenamefont {Yi}, \citenamefont {Zhang}, \citenamefont {Henderson},
  \citenamefont {Chang}, \citenamefont {Korzekwa}, \citenamefont {Tsai} \emph
  {et~al.}}]{Wang2013}%
  \BibitemOpen
  \bibfield  {author} {\bibinfo {author} {\bibfnamefont {X.}~\bibnamefont
  {Wang}}, \bibinfo {author} {\bibfnamefont {R.}~\bibnamefont {Zgadzaj}},
  \bibinfo {author} {\bibfnamefont {N.}~\bibnamefont {Fazel}}, \bibinfo
  {author} {\bibfnamefont {Z.}~\bibnamefont {Li}}, \bibinfo {author}
  {\bibfnamefont {S.~A.}\ \bibnamefont {Yi}}, \bibinfo {author} {\bibfnamefont
  {X.}~\bibnamefont {Zhang}}, \bibinfo {author} {\bibfnamefont
  {W.}~\bibnamefont {Henderson}}, \bibinfo {author} {\bibfnamefont {Y.~Y.}\
  \bibnamefont {Chang}}, \bibinfo {author} {\bibfnamefont {R.}~\bibnamefont
  {Korzekwa}}, \bibinfo {author} {\bibfnamefont {H.-E.}\ \bibnamefont {Tsai}},
  \emph {et~al.},\ }\bibfield  {title} {\enquote {\bibinfo {title}
  {{Quasi-monoenergetic laser-plasma acceleration of electrons to 2~GeV}},}\
  }\href@noop {} {\bibfield  {journal} {\bibinfo  {journal} {Nature
  Communications},\ }\textbf {\bibinfo {volume} {4}},\ \bibinfo {pages} {1988}
  (\bibinfo {year} {2013})}\BibitemShut {NoStop}%
\bibitem [{\citenamefont {Ta~Phuoc}\ \emph {et~al.}(2012)\citenamefont
  {Ta~Phuoc}, \citenamefont {Corde}, \citenamefont {Thaury}, \citenamefont
  {Malka}, \citenamefont {Tafzi}, \citenamefont {Goddet}, \citenamefont {Shah},
  \citenamefont {Sebban},\ and\ \citenamefont {Rousse}}]{TaPhuoc2012a}%
  \BibitemOpen
  \bibfield  {author} {\bibinfo {author} {\bibfnamefont {K.}~\bibnamefont
  {Ta~Phuoc}}, \bibinfo {author} {\bibfnamefont {S.}~\bibnamefont {Corde}},
  \bibinfo {author} {\bibfnamefont {C.}~\bibnamefont {Thaury}}, \bibinfo
  {author} {\bibfnamefont {V.}~\bibnamefont {Malka}}, \bibinfo {author}
  {\bibfnamefont {A.}~\bibnamefont {Tafzi}}, \bibinfo {author} {\bibfnamefont
  {J.~P.}\ \bibnamefont {Goddet}}, \bibinfo {author} {\bibfnamefont {R.~C.}\
  \bibnamefont {Shah}}, \bibinfo {author} {\bibfnamefont {S.}~\bibnamefont
  {Sebban}}, \ and\ \bibinfo {author} {\bibfnamefont {A.}~\bibnamefont
  {Rousse}},\ }\bibfield  {title} {\enquote {\bibinfo {title} {{All-optical
  Compton gamma-ray source}},}\ }\Doi {10.1038/nphoton.2012.82} {\bibfield
  {journal} {\bibinfo  {journal} {Nat. Photon.},\ }\textbf {\bibinfo {volume}
  {6}},\ \bibinfo {pages} {308} (\bibinfo {year} {2012})}\BibitemShut {NoStop}%
\bibitem [{\citenamefont {Chen}\ \emph {et~al.}(2013)\citenamefont {Chen},
  \citenamefont {Powers}, \citenamefont {Ghebregziabher}, \citenamefont
  {Maharjan}, \citenamefont {Liu}, \citenamefont {Golovin}, \citenamefont
  {Banerjee}, \citenamefont {Zhang}, \citenamefont {Cunningham}, \citenamefont
  {Moorti} \emph {et~al.}}]{Chen2013a}%
  \BibitemOpen
  \bibfield  {author} {\bibinfo {author} {\bibfnamefont {S.}~\bibnamefont
  {Chen}}, \bibinfo {author} {\bibfnamefont {N.~D.}\ \bibnamefont {Powers}},
  \bibinfo {author} {\bibfnamefont {I.}~\bibnamefont {Ghebregziabher}},
  \bibinfo {author} {\bibfnamefont {C.~M.}\ \bibnamefont {Maharjan}}, \bibinfo
  {author} {\bibfnamefont {C.}~\bibnamefont {Liu}}, \bibinfo {author}
  {\bibfnamefont {G.}~\bibnamefont {Golovin}}, \bibinfo {author} {\bibfnamefont
  {S.}~\bibnamefont {Banerjee}}, \bibinfo {author} {\bibfnamefont
  {J.}~\bibnamefont {Zhang}}, \bibinfo {author} {\bibfnamefont
  {N.}~\bibnamefont {Cunningham}}, \bibinfo {author} {\bibfnamefont
  {A.}~\bibnamefont {Moorti}},  \emph {et~al.},\ }\bibfield  {title} {\enquote
  {\bibinfo {title} {{MeV-energy X rays from inverse compton scattering with
  laser-wakefield accelerated electrons}},}\ }\Doi
  {10.1103/PhysRevLett.110.155003} {\bibfield  {journal} {\bibinfo  {journal}
  {Phys. Rev. Lett.},\ }\textbf {\bibinfo {volume} {110}},\ \bibinfo {pages}
  {155003} (\bibinfo {year} {2013})}\BibitemShut {NoStop}%
\bibitem [{\citenamefont {Sarri}\ \emph {et~al.}(2014)\citenamefont {Sarri},
  \citenamefont {Corvan}, \citenamefont {Schumaker}, \citenamefont {Cole},
  \citenamefont {Di~Piazza}, \citenamefont {Ahmed}, \citenamefont {Harvey},
  \citenamefont {Keitel}, \citenamefont {Krushelnick}, \citenamefont {Mangles},
  \citenamefont {Najmudin} \emph {et~al.}}]{Sarri2014}%
  \BibitemOpen
  \bibfield  {author} {\bibinfo {author} {\bibfnamefont {G.}~\bibnamefont
  {Sarri}}, \bibinfo {author} {\bibfnamefont {D.~J.}\ \bibnamefont {Corvan}},
  \bibinfo {author} {\bibfnamefont {W.}~\bibnamefont {Schumaker}}, \bibinfo
  {author} {\bibfnamefont {J.~M.}\ \bibnamefont {Cole}}, \bibinfo {author}
  {\bibfnamefont {A.}~\bibnamefont {Di~Piazza}}, \bibinfo {author}
  {\bibfnamefont {H.}~\bibnamefont {Ahmed}}, \bibinfo {author} {\bibfnamefont
  {C.}~\bibnamefont {Harvey}}, \bibinfo {author} {\bibfnamefont {C.~H.}\
  \bibnamefont {Keitel}}, \bibinfo {author} {\bibfnamefont {K.}~\bibnamefont
  {Krushelnick}}, \bibinfo {author} {\bibfnamefont {S.~P.~D.}\ \bibnamefont
  {Mangles}}, \bibinfo {author} {\bibfnamefont {Z.}~\bibnamefont {Najmudin}},
  \emph {et~al.},\ }\bibfield  {title} {\enquote {\bibinfo {title} {{Ultrahigh
  Brilliance Multi-MeV {$\gamma$}-Ray Beams from Nonlinear Relativistic Thomson
  Scattering}},}\ }\Doi {10.1103/PhysRevLett.113.224801} {\bibfield  {journal}
  {\bibinfo  {journal} {Phys. Rev. Lett.},\ }\textbf {\bibinfo {volume}
  {113}},\ \bibinfo {pages} {224801} (\bibinfo {year} {2014})}\BibitemShut
  {NoStop}%
\bibitem [{\citenamefont {Yan}\ \emph {et~al.}(2017)\citenamefont {Yan},
  \citenamefont {Fruhling}, \citenamefont {Golovin}, \citenamefont {Haden},
  \citenamefont {Luo}, \citenamefont {Zhang}, \citenamefont {Zhao},
  \citenamefont {Zhang}, \citenamefont {Liu}, \citenamefont {Chen} \emph
  {et~al.}}]{Yan2017}%
  \BibitemOpen
  \bibfield  {author} {\bibinfo {author} {\bibfnamefont {W.}~\bibnamefont
  {Yan}}, \bibinfo {author} {\bibfnamefont {C.}~\bibnamefont {Fruhling}},
  \bibinfo {author} {\bibfnamefont {G.}~\bibnamefont {Golovin}}, \bibinfo
  {author} {\bibfnamefont {D.}~\bibnamefont {Haden}}, \bibinfo {author}
  {\bibfnamefont {J.}~\bibnamefont {Luo}}, \bibinfo {author} {\bibfnamefont
  {P.}~\bibnamefont {Zhang}}, \bibinfo {author} {\bibfnamefont
  {B.}~\bibnamefont {Zhao}}, \bibinfo {author} {\bibfnamefont {J.}~\bibnamefont
  {Zhang}}, \bibinfo {author} {\bibfnamefont {C.}~\bibnamefont {Liu}}, \bibinfo
  {author} {\bibfnamefont {M.}~\bibnamefont {Chen}},  \emph {et~al.},\
  }\bibfield  {title} {\enquote {\bibinfo {title} {{High-order multiphoton
  Thomson scattering}},}\ }\Doi {10.1038/nphoton.2017.100} {\bibfield
  {journal} {\bibinfo  {journal} {Nat. Photon.},\ \bibinfo {pages} {514Ð520}}
  (\bibinfo {year} {2017})}\BibitemShut {NoStop}%
\bibitem [{\citenamefont {Powers}\ \emph {et~al.}(2014)\citenamefont {Powers},
  \citenamefont {Ghebregziabher}, \citenamefont {Golovin}, \citenamefont {Liu},
  \citenamefont {Chen}, \citenamefont {Banerjee}, \citenamefont {Zhang},\ and\
  \citenamefont {Umstadter}}]{Powers2014}%
  \BibitemOpen
  \bibfield  {author} {\bibinfo {author} {\bibfnamefont {N.~D.}\ \bibnamefont
  {Powers}}, \bibinfo {author} {\bibfnamefont {I.}~\bibnamefont
  {Ghebregziabher}}, \bibinfo {author} {\bibfnamefont {G.}~\bibnamefont
  {Golovin}}, \bibinfo {author} {\bibfnamefont {C.}~\bibnamefont {Liu}},
  \bibinfo {author} {\bibfnamefont {S.}~\bibnamefont {Chen}}, \bibinfo {author}
  {\bibfnamefont {S.}~\bibnamefont {Banerjee}}, \bibinfo {author}
  {\bibfnamefont {J.}~\bibnamefont {Zhang}}, \ and\ \bibinfo {author}
  {\bibfnamefont {D.~P.}\ \bibnamefont {Umstadter}},\ }\bibfield  {title}
  {\enquote {\bibinfo {title} {{Quasi-monoenergetic and tunable X-rays from a
  laser-driven Compton light source}},}\ }\Doi {10.1038/nphoton.2013.314}
  {\bibfield  {journal} {\bibinfo  {journal} {Nat. Photon.},\ }\textbf
  {\bibinfo {volume} {8}},\ \bibinfo {pages} {28} (\bibinfo {year}
  {2014})}\BibitemShut {NoStop}%
\bibitem [{\citenamefont {Tsai}\ \emph {et~al.}(2015)\citenamefont {Tsai},
  \citenamefont {Wang}, \citenamefont {Shaw}, \citenamefont {Arefiev},
  \citenamefont {Li}, \citenamefont {Zhang}, \citenamefont {Zgadzaj},
  \citenamefont {Henderson}, \citenamefont {Khudik}, \citenamefont {Shvets},\
  and\ \citenamefont {Downer}}]{Tsai2015}%
  \BibitemOpen
  \bibfield  {author} {\bibinfo {author} {\bibfnamefont {Hai-En}\ \bibnamefont
  {Tsai}}, \bibinfo {author} {\bibfnamefont {Xiaoming}\ \bibnamefont {Wang}},
  \bibinfo {author} {\bibfnamefont {Joseph}\ \bibnamefont {Shaw}}, \bibinfo
  {author} {\bibfnamefont {Alexey~V.}\ \bibnamefont {Arefiev}}, \bibinfo
  {author} {\bibfnamefont {Zhengyan}\ \bibnamefont {Li}}, \bibinfo {author}
  {\bibfnamefont {Xi}~\bibnamefont {Zhang}}, \bibinfo {author} {\bibfnamefont
  {Rafal}\ \bibnamefont {Zgadzaj}}, \bibinfo {author} {\bibfnamefont {Watson}\
  \bibnamefont {Henderson}}, \bibinfo {author} {\bibfnamefont {V.}~\bibnamefont
  {Khudik}}, \bibinfo {author} {\bibfnamefont {G.}~\bibnamefont {Shvets}}, \
  and\ \bibinfo {author} {\bibfnamefont {M.~C.}\ \bibnamefont {Downer}},\
  }\bibfield  {title} {\enquote {\bibinfo {title} {{Compact tunable Compton
  x-ray source from laser wakefield accelerator and plasma mirror}},}\
  }\href@noop {} {\bibfield  {journal} {\bibinfo  {journal} {Phys. Plasmas},\
  }\textbf {\bibinfo {volume} {22}},\ \bibinfo {pages} {023106} (\bibinfo
  {year} {2015})}\BibitemShut {NoStop}%
\bibitem [{\citenamefont {Sakai}\ \emph {et~al.}(2015)\citenamefont {Sakai},
  \citenamefont {Pogorelsky}, \citenamefont {Williams}, \citenamefont {O'Shea},
  \citenamefont {Barber}, \citenamefont {Gadjev}, \citenamefont {Duris},
  \citenamefont {Musumeci}, \citenamefont {Fedurin}, \citenamefont
  {Korostyshevsky} \emph {et~al.}}]{Sakai2015}%
  \BibitemOpen
  \bibfield  {author} {\bibinfo {author} {\bibfnamefont {Y.}~\bibnamefont
  {Sakai}}, \bibinfo {author} {\bibfnamefont {I.}~\bibnamefont {Pogorelsky}},
  \bibinfo {author} {\bibfnamefont {O.}~\bibnamefont {Williams}}, \bibinfo
  {author} {\bibfnamefont {F.}~\bibnamefont {O'Shea}}, \bibinfo {author}
  {\bibfnamefont {S.}~\bibnamefont {Barber}}, \bibinfo {author} {\bibfnamefont
  {I.}~\bibnamefont {Gadjev}}, \bibinfo {author} {\bibfnamefont
  {J.}~\bibnamefont {Duris}}, \bibinfo {author} {\bibfnamefont
  {P.}~\bibnamefont {Musumeci}}, \bibinfo {author} {\bibfnamefont
  {M.}~\bibnamefont {Fedurin}}, \bibinfo {author} {\bibfnamefont
  {A.}~\bibnamefont {Korostyshevsky}},  \emph {et~al.},\ }\bibfield  {title}
  {\enquote {\bibinfo {title} {{Observation of redshifting and harmonic
  radiation in inverse Compton scattering}},}\ }\Doi
  {10.1103/PhysRevSTAB.18.060702} {\bibfield  {journal} {\bibinfo  {journal}
  {Phys. Rev. Spec. Top. Ac.},\ }\textbf {\bibinfo {volume} {18}},\ \bibinfo
  {pages} {060702} (\bibinfo {year} {2015})}\BibitemShut {NoStop}%
\bibitem [{\citenamefont {Khrennikov}\ \emph {et~al.}(2015)\citenamefont
  {Khrennikov}, \citenamefont {Wenz}, \citenamefont {Buck}, \citenamefont {Xu},
  \citenamefont {Heigoldt}, \citenamefont {Veisz},\ and\ \citenamefont
  {Karsch}}]{Khrennikov2015}%
  \BibitemOpen
  \bibfield  {author} {\bibinfo {author} {\bibfnamefont {K.}~\bibnamefont
  {Khrennikov}}, \bibinfo {author} {\bibfnamefont {J.}~\bibnamefont {Wenz}},
  \bibinfo {author} {\bibfnamefont {A.}~\bibnamefont {Buck}}, \bibinfo {author}
  {\bibfnamefont {J.}~\bibnamefont {Xu}}, \bibinfo {author} {\bibfnamefont
  {M.}~\bibnamefont {Heigoldt}}, \bibinfo {author} {\bibfnamefont
  {L.}~\bibnamefont {Veisz}}, \ and\ \bibinfo {author} {\bibfnamefont
  {S.}~\bibnamefont {Karsch}},\ }\bibfield  {title} {\enquote {\bibinfo {title}
  {{Tunable All-Optical Quasimonochromatic Thomson X-Ray Source in the
  Nonlinear Regime}},}\ }\Doi {10.1103/PhysRevLett.114.195003} {\bibfield
  {journal} {\bibinfo  {journal} {Phys. Rev. Lett.},\ }\textbf {\bibinfo
  {volume} {114}},\ \bibinfo {pages} {195003} (\bibinfo {year}
  {2015})}\BibitemShut {NoStop}%
\bibitem [{\citenamefont {Bula}\ \emph {et~al.}(1996)\citenamefont {Bula},
  \citenamefont {McDonald}, \citenamefont {Prebys}, \citenamefont {Bamber},
  \citenamefont {Boege}, \citenamefont {Kotseroglou}, \citenamefont
  {Melissinos}, \citenamefont {Meyerhofer}, \citenamefont {Ragg}, \citenamefont
  {Burke} \emph {et~al.}}]{Bula1996}%
  \BibitemOpen
  \bibfield  {author} {\bibinfo {author} {\bibfnamefont {C.}~\bibnamefont
  {Bula}}, \bibinfo {author} {\bibfnamefont {K.}~\bibnamefont {McDonald}},
  \bibinfo {author} {\bibfnamefont {E.}~\bibnamefont {Prebys}}, \bibinfo
  {author} {\bibfnamefont {C.}~\bibnamefont {Bamber}}, \bibinfo {author}
  {\bibfnamefont {S.}~\bibnamefont {Boege}}, \bibinfo {author} {\bibfnamefont
  {T.}~\bibnamefont {Kotseroglou}}, \bibinfo {author} {\bibfnamefont
  {A.}~\bibnamefont {Melissinos}}, \bibinfo {author} {\bibfnamefont
  {D.}~\bibnamefont {Meyerhofer}}, \bibinfo {author} {\bibfnamefont
  {W.}~\bibnamefont {Ragg}}, \bibinfo {author} {\bibfnamefont {D.}~\bibnamefont
  {Burke}},  \emph {et~al.},\ }\bibfield  {title} {\enquote {\bibinfo {title}
  {{Observation of Nonlinear Effects in Compton Scattering}},}\ }\Doi
  {10.1103/PhysRevLett.76.3116} {\bibfield  {journal} {\bibinfo  {journal}
  {Phys. Rev. Lett.},\ }\textbf {\bibinfo {volume} {76}},\ \bibinfo {pages}
  {3116} (\bibinfo {year} {1996})}\BibitemShut {NoStop}%
\bibitem [{\citenamefont {Esarey}\ \emph {et~al.}(1993)\citenamefont {Esarey},
  \citenamefont {Ride},\ and\ \citenamefont {Sprangle}}]{Esarey1993b}%
  \BibitemOpen
  \bibfield  {author} {\bibinfo {author} {\bibfnamefont {E.}~\bibnamefont
  {Esarey}}, \bibinfo {author} {\bibfnamefont {S.K.}\ \bibnamefont {Ride}}, \
  and\ \bibinfo {author} {\bibfnamefont {P.}~\bibnamefont {Sprangle}},\
  }\bibfield  {title} {\enquote {\bibinfo {title} {{Nonlinear Thomson
  scattering of intense laser pulses from beams and plasmas}},}\ }\Doi
  {10.1103/PhysRevE.48.3003} {\bibfield  {journal} {\bibinfo  {journal} {Phys.
  Rev. E},\ }\textbf {\bibinfo {volume} {48}},\ \bibinfo {pages} {3003}
  (\bibinfo {year} {1993})}\BibitemShut {NoStop}%
\bibitem [{\citenamefont {Ritus}(1985)}]{Ritus1985a}%
  \BibitemOpen
  \bibfield  {author} {\bibinfo {author} {\bibfnamefont {V.~I.}\ \bibnamefont
  {Ritus}},\ }\bibfield  {title} {\enquote {\bibinfo {title} {{Quantum effects
  of the interaction of elementary particles with an intense electromagnetic
  field}},}\ }\Doi {10.1007/BF01120220} {\bibfield  {journal} {\bibinfo
  {journal} {J. Russ. Laser Res.},\ }\textbf {\bibinfo {volume} {6}},\ \bibinfo
  {pages} {497} (\bibinfo {year} {1985})}\BibitemShut {NoStop}%
\bibitem [{\citenamefont {Nerush}\ and\ \citenamefont
  {Kostyukov}(2011)}]{Nerush2011}%
  \BibitemOpen
  \bibfield  {author} {\bibinfo {author} {\bibfnamefont {E.~N.}\ \bibnamefont
  {Nerush}}\ and\ \bibinfo {author} {\bibfnamefont {I.~Y.}\ \bibnamefont
  {Kostyukov}},\ }\bibfield  {title} {\enquote {\bibinfo {title} {{Kinetic
  modelling of quantum effects in laser-beam interaction}},}\ }\Doi
  {10.1016/j.nima.2011.02.065} {\bibfield  {journal} {\bibinfo  {journal}
  {Nucl. Instrum. Methods A},\ }\textbf {\bibinfo {volume} {653}},\ \bibinfo
  {pages} {7} (\bibinfo {year} {2011})}\BibitemShut {NoStop}%
\bibitem [{\citenamefont {Blackburn}\ \emph {et~al.}(2014)\citenamefont
  {Blackburn}, \citenamefont {Ridgers}, \citenamefont {Kirk},\ and\
  \citenamefont {Bell}}]{Blackburn2014}%
  \BibitemOpen
  \bibfield  {author} {\bibinfo {author} {\bibfnamefont {T.~G.}\ \bibnamefont
  {Blackburn}}, \bibinfo {author} {\bibfnamefont {C.~P.}\ \bibnamefont
  {Ridgers}}, \bibinfo {author} {\bibfnamefont {J.~G.}\ \bibnamefont {Kirk}}, \
  and\ \bibinfo {author} {\bibfnamefont {A.~R.}\ \bibnamefont {Bell}},\
  }\bibfield  {title} {\enquote {\bibinfo {title} {{Quantum Radiation Reaction
  in Laser-Electron-Beam Collisions}},}\ }\Doi {10.1103/PhysRevLett.112.015001}
  {\bibfield  {journal} {\bibinfo  {journal} {Phys. Rev. Lett.},\ }\textbf
  {\bibinfo {volume} {112}},\ \bibinfo {pages} {015001} (\bibinfo {year}
  {2014})}\BibitemShut {NoStop}%
\bibitem [{\citenamefont {Faure}\ \emph {et~al.}(2006)\citenamefont {Faure},
  \citenamefont {Rechatin}, \citenamefont {Norlin}, \citenamefont {Lifschitz},
  \citenamefont {Glinec},\ and\ \citenamefont {Malka}}]{Faure2006}%
  \BibitemOpen
  \bibfield  {author} {\bibinfo {author} {\bibfnamefont {J.}~\bibnamefont
  {Faure}}, \bibinfo {author} {\bibfnamefont {C.}~\bibnamefont {Rechatin}},
  \bibinfo {author} {\bibfnamefont {A.}~\bibnamefont {Norlin}}, \bibinfo
  {author} {\bibfnamefont {A.}~\bibnamefont {Lifschitz}}, \bibinfo {author}
  {\bibfnamefont {Y.}~\bibnamefont {Glinec}}, \ and\ \bibinfo {author}
  {\bibfnamefont {V.}~\bibnamefont {Malka}},\ }\bibfield  {title} {\enquote
  {\bibinfo {title} {{Controlled injection and acceleration of electrons in
  plasma wakefields by colliding laser pulses.}}}\ }\href
  {http://www.ncbi.nlm.nih.gov/pubmed/17151663} {\bibfield  {journal} {\bibinfo
   {journal} {Nature},\ }\textbf {\bibinfo {volume} {444}},\ \bibinfo {pages}
  {737} (\bibinfo {year} {2006})}\BibitemShut {NoStop}%
\bibitem [{\citenamefont {Bulanov}\ \emph {et~al.}(1997)\citenamefont
  {Bulanov}, \citenamefont {Pegoraro}, \citenamefont {Pukhov},\ and\
  \citenamefont {Sakharov}}]{Bulanov1997}%
  \BibitemOpen
  \bibfield  {author} {\bibinfo {author} {\bibfnamefont {S.~V.}\ \bibnamefont
  {Bulanov}}, \bibinfo {author} {\bibfnamefont {F.}~\bibnamefont {Pegoraro}},
  \bibinfo {author} {\bibfnamefont {A.}~\bibnamefont {Pukhov}}, \ and\ \bibinfo
  {author} {\bibfnamefont {A.~S.}\ \bibnamefont {Sakharov}},\ }\bibfield
  {title} {\enquote {\bibinfo {title} {{Transverse-Wake Wave Breaking}},}\
  }\Doi {10.1103/PhysRevLett.78.4205} {\bibfield  {journal} {\bibinfo
  {journal} {Phys. Rev. Lett.},\ }\textbf {\bibinfo {volume} {78}},\ \bibinfo
  {pages} {4205} (\bibinfo {year} {1997})}\BibitemShut {NoStop}%
\bibitem [{\citenamefont {Schmid}\ \emph {et~al.}(2010)\citenamefont {Schmid},
  \citenamefont {Buck}, \citenamefont {Sears}, \citenamefont {Mikhailova},
  \citenamefont {Tautz}, \citenamefont {Herrmann}, \citenamefont {Geissler},
  \citenamefont {Krausz},\ and\ \citenamefont {Veisz}}]{Schmid2010}%
  \BibitemOpen
  \bibfield  {author} {\bibinfo {author} {\bibfnamefont {K.}~\bibnamefont
  {Schmid}}, \bibinfo {author} {\bibfnamefont {A}~\bibnamefont {Buck}},
  \bibinfo {author} {\bibfnamefont {C.~M.~S.}\ \bibnamefont {Sears}}, \bibinfo
  {author} {\bibfnamefont {J~M}\ \bibnamefont {Mikhailova}}, \bibinfo {author}
  {\bibfnamefont {R}~\bibnamefont {Tautz}}, \bibinfo {author} {\bibfnamefont
  {D}~\bibnamefont {Herrmann}}, \bibinfo {author} {\bibfnamefont
  {M}~\bibnamefont {Geissler}}, \bibinfo {author} {\bibfnamefont
  {F}~\bibnamefont {Krausz}}, \ and\ \bibinfo {author} {\bibfnamefont
  {L.}~\bibnamefont {Veisz}},\ }\bibfield  {title} {\enquote {\bibinfo {title}
  {{Density-transition based electron injector for laser driven wakefield
  accelerators}},}\ }\href
  {http://link.aps.org/doi/10.1103/PhysRevSTAB.13.091301} {\bibfield  {journal}
  {\bibinfo  {journal} {Physical Review Special Topics - Accelerators and
  Beams},\ }\textbf {\bibinfo {volume} {13}},\ \bibinfo {pages} {91301}
  (\bibinfo {year} {2010})}\BibitemShut {NoStop}%
\bibitem [{\citenamefont {Ridgers}\ \emph {et~al.}(2014)\citenamefont
  {Ridgers}, \citenamefont {Kirk}, \citenamefont {Duclous}, \citenamefont
  {Blackburn}, \citenamefont {Brady}, \citenamefont {Bennett}, \citenamefont
  {Arber},\ and\ \citenamefont {Bell}}]{Ridgers2014}%
  \BibitemOpen
  \bibfield  {author} {\bibinfo {author} {\bibfnamefont {C.~P.}\ \bibnamefont
  {Ridgers}}, \bibinfo {author} {\bibfnamefont {J.~G.}\ \bibnamefont {Kirk}},
  \bibinfo {author} {\bibfnamefont {R.}~\bibnamefont {Duclous}}, \bibinfo
  {author} {\bibfnamefont {T.~G.}\ \bibnamefont {Blackburn}}, \bibinfo {author}
  {\bibfnamefont {C.S.}\ \bibnamefont {Brady}}, \bibinfo {author}
  {\bibfnamefont {K.}~\bibnamefont {Bennett}}, \bibinfo {author} {\bibfnamefont
  {T.~D.}\ \bibnamefont {Arber}}, \ and\ \bibinfo {author} {\bibfnamefont
  {A.~R.}\ \bibnamefont {Bell}},\ }\bibfield  {title} {\enquote {\bibinfo
  {title} {{Modelling gamma-ray photon emission and pair production in
  high-intensity laser-matter interactions}},}\ }\Doi
  {10.1016/j.jcp.2013.12.007} {\bibfield  {journal} {\bibinfo  {journal} {J.
  Comput. Phys.},\ }\textbf {\bibinfo {volume} {260}},\ \bibinfo {pages} {273}
  (\bibinfo {year} {2014})}\BibitemShut {NoStop}%
\bibitem [{\citenamefont {Decker}\ \emph {et~al.}(1996)\citenamefont {Decker},
  \citenamefont {Mori}, \citenamefont {Tzeng},\ and\ \citenamefont
  {Katsouleas}}]{Decker1996a}%
  \BibitemOpen
  \bibfield  {author} {\bibinfo {author} {\bibfnamefont {C.~D.}\ \bibnamefont
  {Decker}}, \bibinfo {author} {\bibfnamefont {W.~B.}\ \bibnamefont {Mori}},
  \bibinfo {author} {\bibfnamefont {K.-C.}\ \bibnamefont {Tzeng}}, \ and\
  \bibinfo {author} {\bibfnamefont {T.}~\bibnamefont {Katsouleas}},\ }\bibfield
   {title} {\enquote {\bibinfo {title} {{The evolution of ultra-intense,
  short-pulse lasers in underdense plasmas}},}\ }\Doi {10.1063/1.872001}
  {\bibfield  {journal} {\bibinfo  {journal} {Phys. Plasmas},\ }\textbf
  {\bibinfo {volume} {3}},\ \bibinfo {pages} {2047} (\bibinfo {year}
  {1996})}\BibitemShut {NoStop}%
\bibitem [{\citenamefont {Agostinelli}\ \emph {et~al.}(2003)\citenamefont
  {Agostinelli}, \citenamefont {Allison}, \citenamefont {Amako}, \citenamefont
  {Apostolakis}, \citenamefont {Araujo}, \citenamefont {Arce}, \citenamefont
  {Asai}, \citenamefont {Axen}, \citenamefont {Banerjee}, \citenamefont
  {Barrand} \emph {et~al.}}]{Agostinelli2003}%
  \BibitemOpen
  \bibfield  {author} {\bibinfo {author} {\bibfnamefont {S.}~\bibnamefont
  {Agostinelli}}, \bibinfo {author} {\bibfnamefont {J.}~\bibnamefont
  {Allison}}, \bibinfo {author} {\bibfnamefont {K.}~\bibnamefont {Amako}},
  \bibinfo {author} {\bibfnamefont {J.}~\bibnamefont {Apostolakis}}, \bibinfo
  {author} {\bibfnamefont {H.}~\bibnamefont {Araujo}}, \bibinfo {author}
  {\bibfnamefont {P.}~\bibnamefont {Arce}}, \bibinfo {author} {\bibfnamefont
  {M.}~\bibnamefont {Asai}}, \bibinfo {author} {\bibfnamefont {D.}~\bibnamefont
  {Axen}}, \bibinfo {author} {\bibfnamefont {S.}~\bibnamefont {Banerjee}},
  \bibinfo {author} {\bibfnamefont {G.}~\bibnamefont {Barrand}},  \emph
  {et~al.},\ }\bibfield  {title} {\enquote {\bibinfo {title} {{GEANT4 - A
  simulation toolkit}},}\ }\Doi {10.1016/S0168-9002(03)01368-8} {\bibfield
  {journal} {\bibinfo  {journal} {Nucl. Instrum. Methods A},\ }\textbf
  {\bibinfo {volume} {506}},\ \bibinfo {pages} {250} (\bibinfo {year}
  {2003})}\BibitemShut {NoStop}%
\bibitem [{\citenamefont {Goorley}\ \emph {et~al.}(2012)\citenamefont
  {Goorley}, \citenamefont {James}, \citenamefont {Booth}, \citenamefont
  {Brown}, \citenamefont {Bull}, \citenamefont {Cox}, \citenamefont {Durkee},
  \citenamefont {Elson}, \citenamefont {Fensin}, \citenamefont {Forster} \emph
  {et~al.}}]{Goorley2012}%
  \BibitemOpen
  \bibfield  {author} {\bibinfo {author} {\bibfnamefont {T.}~\bibnamefont
  {Goorley}}, \bibinfo {author} {\bibfnamefont {M.}~\bibnamefont {James}},
  \bibinfo {author} {\bibfnamefont {T.}~\bibnamefont {Booth}}, \bibinfo
  {author} {\bibfnamefont {F.}~\bibnamefont {Brown}}, \bibinfo {author}
  {\bibfnamefont {J.}~\bibnamefont {Bull}}, \bibinfo {author} {\bibfnamefont
  {L.~J.}\ \bibnamefont {Cox}}, \bibinfo {author} {\bibfnamefont
  {J.}~\bibnamefont {Durkee}}, \bibinfo {author} {\bibfnamefont
  {J.}~\bibnamefont {Elson}}, \bibinfo {author} {\bibfnamefont
  {M.}~\bibnamefont {Fensin}}, \bibinfo {author} {\bibfnamefont {R.~A.}\
  \bibnamefont {Forster}},  \emph {et~al.},\ }\bibfield  {title} {\enquote
  {\bibinfo {title} {{Initial MCNP6 Release Overview}},}\ }\Doi
  {10.13182/NT11-135} {\bibfield  {journal} {\bibinfo  {journal} {Nucl.
  Technol.},\ }\textbf {\bibinfo {volume} {180}},\ \bibinfo {pages} {298}
  (\bibinfo {year} {2012})}\BibitemShut {NoStop}%
\bibitem [{\citenamefont {Frle{\v{z}}}\ \emph {et~al.}(2000)\citenamefont
  {Frle{\v{z}}}, \citenamefont {Supek}, \citenamefont {Assamagan},
  \citenamefont {Br{\"{o}}nnimann}, \citenamefont {Fl{\"{u}}gel}, \citenamefont
  {Krause}, \citenamefont {Lawrence}, \citenamefont {Mzavia}, \citenamefont
  {Po{\v{c}}ani{\'{c}}}, \citenamefont {Renker} \emph {et~al.}}]{Frlez2000}%
  \BibitemOpen
  \bibfield  {author} {\bibinfo {author} {\bibfnamefont {E.}~\bibnamefont
  {Frle{\v{z}}}}, \bibinfo {author} {\bibfnamefont {I.}~\bibnamefont {Supek}},
  \bibinfo {author} {\bibfnamefont {K.~A.}\ \bibnamefont {Assamagan}}, \bibinfo
  {author} {\bibfnamefont {C.}~\bibnamefont {Br{\"{o}}nnimann}}, \bibinfo
  {author} {\bibfnamefont {T.}~\bibnamefont {Fl{\"{u}}gel}}, \bibinfo {author}
  {\bibfnamefont {B.}~\bibnamefont {Krause}}, \bibinfo {author} {\bibfnamefont
  {D.~W.}\ \bibnamefont {Lawrence}}, \bibinfo {author} {\bibfnamefont
  {D.}~\bibnamefont {Mzavia}}, \bibinfo {author} {\bibfnamefont
  {D.}~\bibnamefont {Po{\v{c}}ani{\'{c}}}}, \bibinfo {author} {\bibfnamefont
  {D.}~\bibnamefont {Renker}},  \emph {et~al.},\ }\bibfield  {title} {\enquote
  {\bibinfo {title} {{Cosmic muon tomography of pure cesium iodide calorimeter
  crystals}},}\ }\Doi {10.1016/S0168-9002(99)00886-4} {\bibfield  {journal}
  {\bibinfo  {journal} {Nucl. Instrum. Methods A},\ }\textbf {\bibinfo {volume}
  {440}},\ \bibinfo {pages} {57} (\bibinfo {year} {2000})}\BibitemShut
  {NoStop}%
\bibitem [{\citenamefont {Behm}\ \emph {et~al.}(2017)\citenamefont {Behm},
  \citenamefont {Cole}, \citenamefont {Wood}, \citenamefont {Gerstmayr},
  \citenamefont {Poder}, \citenamefont {Mangles}, \citenamefont {Najmudin},
  \citenamefont {Thomas}, \citenamefont {Krushelnick}, \citenamefont {Murphy}
  \emph {et~al.}}]{Behm2017}%
  \BibitemOpen
  \bibfield  {author} {\bibinfo {author} {\bibfnamefont {K.~T.}\ \bibnamefont
  {Behm}}, \bibinfo {author} {\bibfnamefont {J.~M.}\ \bibnamefont {Cole}},
  \bibinfo {author} {\bibfnamefont {J.~C.}\ \bibnamefont {Wood}}, \bibinfo
  {author} {\bibfnamefont {E.}~\bibnamefont {Gerstmayr}}, \bibinfo {author}
  {\bibfnamefont {K.}~\bibnamefont {Poder}}, \bibinfo {author} {\bibfnamefont
  {S.~P.~D.}\ \bibnamefont {Mangles}}, \bibinfo {author} {\bibfnamefont
  {Z.}~\bibnamefont {Najmudin}}, \bibinfo {author} {\bibfnamefont {A.~G.~R.}\
  \bibnamefont {Thomas}}, \bibinfo {author} {\bibfnamefont {K.}~\bibnamefont
  {Krushelnick}}, \bibinfo {author} {\bibfnamefont {C.~D.}\ \bibnamefont
  {Murphy}},  \emph {et~al.},\ }\bibfield  {title} {\enquote {\bibinfo {title}
  {{Novel design of a gamma ray spectrometer measuring spectra with photon
  energies greater than 100 MeV}},}\ }\href@noop {} {\bibfield  {journal}
  {\bibinfo  {journal} {Submitted}} (\bibinfo {year} {2017})}\BibitemShut
  {NoStop}%
\bibitem [{\citenamefont {Harvey}\ \emph {et~al.}(2015)\citenamefont {Harvey},
  \citenamefont {Ilderton},\ and\ \citenamefont {King}}]{Harvey2015}%
  \BibitemOpen
  \bibfield  {author} {\bibinfo {author} {\bibfnamefont {C.~N.}\ \bibnamefont
  {Harvey}}, \bibinfo {author} {\bibfnamefont {A.}~\bibnamefont {Ilderton}}, \
  and\ \bibinfo {author} {\bibfnamefont {B.}~\bibnamefont {King}},\ }\bibfield
  {title} {\enquote {\bibinfo {title} {{Testing numerical implementations of
  strong-field electrodynamics}},}\ }\Doi {10.1103/PhysRevA.91.013822}
  {\bibfield  {journal} {\bibinfo  {journal} {Phys. Rev. A},\ }\textbf
  {\bibinfo {volume} {91}},\ \bibinfo {pages} {013822} (\bibinfo {year}
  {2015})}\BibitemShut {NoStop}%
\bibitem [{\citenamefont {Gonoskov}\ \emph {et~al.}(2015)\citenamefont
  {Gonoskov}, \citenamefont {Bastrakov}, \citenamefont {Efimenko},
  \citenamefont {Ilderton}, \citenamefont {Marklund}, \citenamefont {Meyerov},
  \citenamefont {Muraviev}, \citenamefont {Sergeev}, \citenamefont {Surmin},\
  and\ \citenamefont {Wallin}}]{Gonoskov2015}%
  \BibitemOpen
  \bibfield  {author} {\bibinfo {author} {\bibfnamefont {A.}~\bibnamefont
  {Gonoskov}}, \bibinfo {author} {\bibfnamefont {S.}~\bibnamefont {Bastrakov}},
  \bibinfo {author} {\bibfnamefont {E.}~\bibnamefont {Efimenko}}, \bibinfo
  {author} {\bibfnamefont {A.}~\bibnamefont {Ilderton}}, \bibinfo {author}
  {\bibfnamefont {M.}~\bibnamefont {Marklund}}, \bibinfo {author}
  {\bibfnamefont {I.}~\bibnamefont {Meyerov}}, \bibinfo {author} {\bibfnamefont
  {A.}~\bibnamefont {Muraviev}}, \bibinfo {author} {\bibfnamefont
  {A.}~\bibnamefont {Sergeev}}, \bibinfo {author} {\bibfnamefont
  {I.}~\bibnamefont {Surmin}}, \ and\ \bibinfo {author} {\bibfnamefont
  {E.}~\bibnamefont {Wallin}},\ }\bibfield  {title} {\enquote {\bibinfo {title}
  {{Extended particle-in-cell schemes for physics in ultrastrong laser fields:
  Review and developments}},}\ }\Doi {10.1103/PhysRevE.92.023305} {\bibfield
  {journal} {\bibinfo  {journal} {Phys. Rev. E},\ }\textbf {\bibinfo {volume}
  {92}},\ \bibinfo {pages} {023305} (\bibinfo {year} {2015})}\BibitemShut
  {NoStop}%
\bibitem [{\citenamefont {Bulanov}\ \emph {et~al.}(2013)\citenamefont
  {Bulanov}, \citenamefont {Schroeder}, \citenamefont {Esarey},\ and\
  \citenamefont {Leemans}}]{Bulanov2013}%
  \BibitemOpen
  \bibfield  {author} {\bibinfo {author} {\bibfnamefont {S.~S.}\ \bibnamefont
  {Bulanov}}, \bibinfo {author} {\bibfnamefont {C.~B.}\ \bibnamefont
  {Schroeder}}, \bibinfo {author} {\bibfnamefont {E.}~\bibnamefont {Esarey}}, \
  and\ \bibinfo {author} {\bibfnamefont {W.~P.}\ \bibnamefont {Leemans}},\
  }\bibfield  {title} {\enquote {\bibinfo {title} {{Electromagnetic cascade in
  high-energy electron, positron, and photon interactions with intense laser
  pulses}},}\ }\Doi {10.1103/PhysRevA.87.062110} {\bibfield  {journal}
  {\bibinfo  {journal} {Phys. Rev. A},\ }\textbf {\bibinfo {volume} {87}},\
  \bibinfo {pages} {062110} (\bibinfo {year} {2013})}\BibitemShut {NoStop}%
\bibitem [{\citenamefont {Sokolov}\ \emph {et~al.}(2010)\citenamefont
  {Sokolov}, \citenamefont {Naumova}, \citenamefont {Nees},\ and\ \citenamefont
  {Mourou}}]{Sokolov2010}%
  \BibitemOpen
  \bibfield  {author} {\bibinfo {author} {\bibfnamefont {I.~V.}\ \bibnamefont
  {Sokolov}}, \bibinfo {author} {\bibfnamefont {N.~M.}\ \bibnamefont
  {Naumova}}, \bibinfo {author} {\bibfnamefont {J.~A.}\ \bibnamefont {Nees}}, \
  and\ \bibinfo {author} {\bibfnamefont {G.A.}\ \bibnamefont {Mourou}},\
  }\bibfield  {title} {\enquote {\bibinfo {title} {{Pair Creation in QED-Strong
  Pulsed Laser Fields Interacting with Electron Beams}},}\ }\Doi
  {10.1103/PhysRevLett.105.195005} {\bibfield  {journal} {\bibinfo  {journal}
  {Phys. Rev. Lett.},\ }\textbf {\bibinfo {volume} {105}},\ \bibinfo {pages}
  {195005} (\bibinfo {year} {2010})}\BibitemShut {NoStop}%
\bibitem [{\citenamefont {Elkina}\ \emph {et~al.}(2011)\citenamefont {Elkina},
  \citenamefont {Fedotov}, \citenamefont {Kostyukov}, \citenamefont {Legkov},
  \citenamefont {Narozhny}, \citenamefont {Nerush},\ and\ \citenamefont
  {Ruhl}}]{Elkina2011}%
  \BibitemOpen
  \bibfield  {author} {\bibinfo {author} {\bibfnamefont {N.~V.}\ \bibnamefont
  {Elkina}}, \bibinfo {author} {\bibfnamefont {A.~M.}\ \bibnamefont {Fedotov}},
  \bibinfo {author} {\bibfnamefont {I.~Yu}\ \bibnamefont {Kostyukov}}, \bibinfo
  {author} {\bibfnamefont {M.~V.}\ \bibnamefont {Legkov}}, \bibinfo {author}
  {\bibfnamefont {N.~B.}\ \bibnamefont {Narozhny}}, \bibinfo {author}
  {\bibfnamefont {E.~N.}\ \bibnamefont {Nerush}}, \ and\ \bibinfo {author}
  {\bibfnamefont {H.}~\bibnamefont {Ruhl}},\ }\bibfield  {title} {\enquote
  {\bibinfo {title} {{QED cascades induced by circularly polarized laser
  fields}},}\ }\Doi {10.1103/PhysRevSTAB.14.054401} {\bibfield  {journal}
  {\bibinfo  {journal} {Phys. Rev. Spec. Top. Ac.},\ }\textbf {\bibinfo
  {volume} {14}},\ \bibinfo {pages} {054401} (\bibinfo {year}
  {2011})}\BibitemShut {NoStop}%
\bibitem [{\citenamefont {Arber}\ \emph {et~al.}(2015)\citenamefont {Arber},
  \citenamefont {Bennett}, \citenamefont {Brady}, \citenamefont {Ramsay},
  \citenamefont {Sircombe}, \citenamefont {Gillies}, \citenamefont {Evans},
  \citenamefont {Schmitz}, \citenamefont {Bell},\ and\ \citenamefont
  {Ridgers}}]{Arber2015}%
  \BibitemOpen
  \bibfield  {author} {\bibinfo {author} {\bibfnamefont {T.~D.}\ \bibnamefont
  {Arber}}, \bibinfo {author} {\bibfnamefont {K.}~\bibnamefont {Bennett}},
  \bibinfo {author} {\bibfnamefont {C.~S.}\ \bibnamefont {Brady}}, \bibinfo
  {author} {\bibfnamefont {M.~G.}\ \bibnamefont {Ramsay}}, \bibinfo {author}
  {\bibfnamefont {N.~J.}\ \bibnamefont {Sircombe}}, \bibinfo {author}
  {\bibfnamefont {P.}~\bibnamefont {Gillies}}, \bibinfo {author} {\bibfnamefont
  {R.~G.}\ \bibnamefont {Evans}}, \bibinfo {author} {\bibfnamefont
  {H.}~\bibnamefont {Schmitz}}, \bibinfo {author} {\bibfnamefont {A.~R.}\
  \bibnamefont {Bell}}, \ and\ \bibinfo {author} {\bibfnamefont {C.~P.}\
  \bibnamefont {Ridgers}},\ }\bibfield  {title} {\enquote {\bibinfo {title}
  {{Contemporary particle-in-cell approach to laser-plasma modelling}},}\ }\Doi
  {10.1088/0741-3335/57/11/113001} {\bibfield  {journal} {\bibinfo  {journal}
  {Plasma Phys. Contr. F.},\ }\textbf {\bibinfo {volume} {57}},\ \bibinfo
  {pages} {113001} (\bibinfo {year} {2015})}\BibitemShut {NoStop}%
\bibitem [{\citenamefont {Green}\ and\ \citenamefont
  {Harvey}(2015)}]{Green2015}%
  \BibitemOpen
  \bibfield  {author} {\bibinfo {author} {\bibfnamefont {D.~G.}\ \bibnamefont
  {Green}}\ and\ \bibinfo {author} {\bibfnamefont {C.~N.}\ \bibnamefont
  {Harvey}},\ }\bibfield  {title} {\enquote {\bibinfo {title} {{SIMLA:
  Simulating particle dynamics in intense laser and other electromagnetic
  fields via classical and quantum electrodynamics}},}\ }\Doi
  {10.1016/j.cpc.2015.02.030} {\bibfield  {journal} {\bibinfo  {journal}
  {Comput. Phys. Commun.},\ }\textbf {\bibinfo {volume} {192}},\ \bibinfo
  {pages} {313} (\bibinfo {year} {2015})}\BibitemShut {NoStop}%
\bibitem [{\citenamefont {Blackburn}(2015)}]{Blackburn2015}%
  \BibitemOpen
  \bibfield  {author} {\bibinfo {author} {\bibfnamefont {T.~G.}\ \bibnamefont
  {Blackburn}},\ }\bibfield  {title} {\enquote {\bibinfo {title} {{Measuring
  quantum radiation reaction in laser-electron-beam collisions}},}\ }\Doi
  {10.1088/0741-3335/57/7/075012} {\bibfield  {journal} {\bibinfo  {journal}
  {Plasma Phys. Contr. F.},\ }\textbf {\bibinfo {volume} {57}},\ \bibinfo
  {pages} {075012} (\bibinfo {year} {2015})}\BibitemShut {NoStop}%
\bibitem [{\citenamefont {Vranic}\ \emph
  {et~al.}(2016){\natexlab{b}}\citenamefont {Vranic}, \citenamefont {Martins},
  \citenamefont {Fonseca},\ and\ \citenamefont {Silva}}]{Vranic2016}%
  \BibitemOpen
  \bibfield  {author} {\bibinfo {author} {\bibfnamefont {M.}~\bibnamefont
  {Vranic}}, \bibinfo {author} {\bibfnamefont {J.~L.}\ \bibnamefont {Martins}},
  \bibinfo {author} {\bibfnamefont {R.~A.}\ \bibnamefont {Fonseca}}, \ and\
  \bibinfo {author} {\bibfnamefont {L.~O.}\ \bibnamefont {Silva}},\ }\bibfield
  {title} {\enquote {\bibinfo {title} {{Classical radiation reaction in
  particle-in-cell simulations}},}\ }\Doi {10.1016/j.cpc.2016.04.002}
  {\bibfield  {journal} {\bibinfo  {journal} {Comput. Phys. Commun.},\ }\textbf
  {\bibinfo {volume} {204}},\ \bibinfo {pages} {141} (\bibinfo {year}
  {2016}{\natexlab{b}})}\BibitemShut {NoStop}%
\bibitem [{\citenamefont {Ridgers}\ \emph {et~al.}(2017)\citenamefont
  {Ridgers}, \citenamefont {Blackburn}, \citenamefont {{Del Sorbo}},
  \citenamefont {Bradley}, \citenamefont {Slade-Lowther}, \citenamefont
  {Baird}, \citenamefont {Mangles}, \citenamefont {McKenna}, \citenamefont
  {Marklund}, \citenamefont {Murphy},\ and\ \citenamefont
  {Thomas}}]{Ridgers2017}%
  \BibitemOpen
  \bibfield  {author} {\bibinfo {author} {\bibfnamefont {C.P.}\ \bibnamefont
  {Ridgers}}, \bibinfo {author} {\bibfnamefont {T.G.}\ \bibnamefont
  {Blackburn}}, \bibinfo {author} {\bibfnamefont {D.}~\bibnamefont {{Del
  Sorbo}}}, \bibinfo {author} {\bibfnamefont {L.E.}\ \bibnamefont {Bradley}},
  \bibinfo {author} {\bibfnamefont {C.}~\bibnamefont {Slade-Lowther}}, \bibinfo
  {author} {\bibfnamefont {C.D.}\ \bibnamefont {Baird}}, \bibinfo {author}
  {\bibfnamefont {S.P.D.}\ \bibnamefont {Mangles}}, \bibinfo {author}
  {\bibfnamefont {P.}~\bibnamefont {McKenna}}, \bibinfo {author} {\bibfnamefont
  {M.}~\bibnamefont {Marklund}}, \bibinfo {author} {\bibfnamefont {C.D.}\
  \bibnamefont {Murphy}}, \ and\ \bibinfo {author} {\bibfnamefont {A.G.R.}\
  \bibnamefont {Thomas}},\ }\bibfield  {title} {\enquote {\bibinfo {title}
  {{Signatures of quantum effects on radiation reaction in laser-electron-beam
  collisions}},}\ }\href@noop {} {\bibfield  {journal} {\bibinfo  {journal}
  {Journal of Plasma Physics},\ }\textbf {\bibinfo {volume} {83}},\ \bibinfo
  {pages} {715830502} (\bibinfo {year} {2017})}\BibitemShut {NoStop}%
\bibitem [{\citenamefont {Schnell}\ \emph {et~al.}(2012)\citenamefont
  {Schnell}, \citenamefont {Savert}, \citenamefont {Landgraf}, \citenamefont
  {Reuter}, \citenamefont {Nicolai}, \citenamefont {Jackel}, \citenamefont
  {Peth}, \citenamefont {Thiele}, \citenamefont {Jansen}, \citenamefont
  {Pukhov} \emph {et~al.}}]{Schnell2012}%
  \BibitemOpen
  \bibfield  {author} {\bibinfo {author} {\bibfnamefont {M.}~\bibnamefont
  {Schnell}}, \bibinfo {author} {\bibfnamefont {A.}~\bibnamefont {Savert}},
  \bibinfo {author} {\bibfnamefont {B.}~\bibnamefont {Landgraf}}, \bibinfo
  {author} {\bibfnamefont {M.}~\bibnamefont {Reuter}}, \bibinfo {author}
  {\bibfnamefont {M.}~\bibnamefont {Nicolai}}, \bibinfo {author} {\bibfnamefont
  {O.}~\bibnamefont {Jackel}}, \bibinfo {author} {\bibfnamefont
  {C.}~\bibnamefont {Peth}}, \bibinfo {author} {\bibfnamefont {T.}~\bibnamefont
  {Thiele}}, \bibinfo {author} {\bibfnamefont {O.}~\bibnamefont {Jansen}},
  \bibinfo {author} {\bibfnamefont {A.}~\bibnamefont {Pukhov}},  \emph
  {et~al.},\ }\bibfield  {title} {\enquote {\bibinfo {title} {{Deducing the
  Electron-Beam Diameter in a Laser-Plasma Accelerator Using X-Ray Betatron
  Radiation}},}\ }\Doi {10.1103/PhysRevLett.108.075001} {\bibfield  {journal}
  {\bibinfo  {journal} {Phys. Rev. Lett.},\ }\textbf {\bibinfo {volume}
  {108}},\ \bibinfo {pages} {075001} (\bibinfo {year} {2012})}\BibitemShut
  {NoStop}%
\bibitem [{\citenamefont {Harvey}\ \emph {et~al.}(2016)\citenamefont {Harvey},
  \citenamefont {Marklund},\ and\ \citenamefont {Holkundkar}}]{Harvey2016}%
  \BibitemOpen
  \bibfield  {author} {\bibinfo {author} {\bibfnamefont {C.}~\bibnamefont
  {Harvey}}, \bibinfo {author} {\bibfnamefont {M.}~\bibnamefont {Marklund}}, \
  and\ \bibinfo {author} {\bibfnamefont {A.~R.}\ \bibnamefont {Holkundkar}},\
  }\bibfield  {title} {\enquote {\bibinfo {title} {{Focusing effects in
  laser-electron Thomson scattering}},}\ }\Doi
  {10.1103/PhysRevAccelBeams.19.094701} {\bibfield  {journal} {\bibinfo
  {journal} {Phys. Rev. Accel. Beams},\ }\textbf {\bibinfo {volume} {19}},\
  \bibinfo {pages} {094701} (\bibinfo {year} {2016})}\BibitemShut {NoStop}%
\bibitem [{\citenamefont {Buck}\ \emph {et~al.}(2013)\citenamefont {Buck},
  \citenamefont {Wenz}, \citenamefont {Xu}, \citenamefont {Khrennikov},
  \citenamefont {Schmid}, \citenamefont {Heigoldt}, \citenamefont {Mikhailova},
  \citenamefont {Geissler}, \citenamefont {Shen}, \citenamefont {Krausz} \emph
  {et~al.}}]{Buck2013}%
  \BibitemOpen
  \bibfield  {author} {\bibinfo {author} {\bibfnamefont {A.}~\bibnamefont
  {Buck}}, \bibinfo {author} {\bibfnamefont {J.}~\bibnamefont {Wenz}}, \bibinfo
  {author} {\bibfnamefont {J.}~\bibnamefont {Xu}}, \bibinfo {author}
  {\bibfnamefont {K.}~\bibnamefont {Khrennikov}}, \bibinfo {author}
  {\bibfnamefont {K.}~\bibnamefont {Schmid}}, \bibinfo {author} {\bibfnamefont
  {M.}~\bibnamefont {Heigoldt}}, \bibinfo {author} {\bibfnamefont {J.~M.}\
  \bibnamefont {Mikhailova}}, \bibinfo {author} {\bibfnamefont
  {M.}~\bibnamefont {Geissler}}, \bibinfo {author} {\bibfnamefont
  {B.}~\bibnamefont {Shen}}, \bibinfo {author} {\bibfnamefont {F.}~\bibnamefont
  {Krausz}},  \emph {et~al.},\ }\bibfield  {title} {\enquote {\bibinfo {title}
  {{Shock-front injector for high-quality laser-plasma acceleration}},}\ }\Doi
  {10.1103/PhysRevLett.110.185006} {\bibfield  {journal} {\bibinfo  {journal}
  {Phys. Rev. Lett.},\ }\textbf {\bibinfo {volume} {110}},\ \bibinfo {pages}
  {185006} (\bibinfo {year} {2013})}\BibitemShut {NoStop}%
\bibitem [{\citenamefont {Gonsalves}\ \emph {et~al.}(2011)\citenamefont
  {Gonsalves}, \citenamefont {Nakamura}, \citenamefont {Lin}, \citenamefont
  {Panasenko}, \citenamefont {Shiraishi}, \citenamefont {Sokollik},
  \citenamefont {Benedetti}, \citenamefont {Schroeder}, \citenamefont {Geddes},
  \citenamefont {van Tilborg} \emph {et~al.}}]{Gonsalves2011}%
  \BibitemOpen
  \bibfield  {author} {\bibinfo {author} {\bibfnamefont {A.~J.}\ \bibnamefont
  {Gonsalves}}, \bibinfo {author} {\bibfnamefont {K.}~\bibnamefont {Nakamura}},
  \bibinfo {author} {\bibfnamefont {C.}~\bibnamefont {Lin}}, \bibinfo {author}
  {\bibfnamefont {D.}~\bibnamefont {Panasenko}}, \bibinfo {author}
  {\bibfnamefont {S.}~\bibnamefont {Shiraishi}}, \bibinfo {author}
  {\bibfnamefont {T.}~\bibnamefont {Sokollik}}, \bibinfo {author}
  {\bibfnamefont {C.}~\bibnamefont {Benedetti}}, \bibinfo {author}
  {\bibfnamefont {C.~B.}\ \bibnamefont {Schroeder}}, \bibinfo {author}
  {\bibfnamefont {C.~G.~R.}\ \bibnamefont {Geddes}}, \bibinfo {author}
  {\bibfnamefont {J.}~\bibnamefont {van Tilborg}},  \emph {et~al.},\ }\bibfield
   {title} {\enquote {\bibinfo {title} {{Tunable laser plasma accelerator based
  on longitudinal density tailoring}},}\ }\Doi {10.1038/nphys2071} {\bibfield
  {journal} {\bibinfo  {journal} {Nature Physics},\ }\textbf {\bibinfo {volume}
  {7}},\ \bibinfo {pages} {862--866} (\bibinfo {year} {2011})}\BibitemShut
  {NoStop}%
\end{thebibliography}%

\end{document}